\documentclass[twocolumn,preprint2]{aastex63}

\usepackage{natbib}      
\usepackage{graphicx}    
\usepackage{amsmath}
\usepackage{times}
\usepackage{amssymb}
\usepackage{stmaryrd} 
\usepackage{color}
\usepackage{bm} 
\usepackage{sidecap}
\sidecaptionvpos{figure}{t}

\usepackage{comment}

\usepackage{fancyhdr}
\usepackage{epsfig}
\usepackage{float}

\usepackage{url}

\usepackage{array}
\usepackage{verbatim}
\usepackage{calc, layouts}

\hypersetup{linkcolor=red,citecolor=cyan,filecolor=cyan,urlcolor=blue}

\received{March 16, 2021}
\revised{May 2, 2021}
\accepted{May 3, 2021}

\submitjournal{ApJS}

\shorttitle{Optimization of \RBSL\ MFRs}
\shortauthors{Titov et al.}

\definecolor{Light}{gray}{.50}
\definecolor{Dark}{gray}{.20}

\definecolor{dark-red}{rgb}{0.8,0,0}
\definecolor{dark-green}{rgb}{0,0.6,0}
\definecolor{dark-blue}{rgb}{0,0,0.8}
\definecolor{dark-margenta}{rgb}{0.8,0,0.8}
\definecolor{dark-purple}{rgb}{0.45,0.2,0.65}
\definecolor{orange}{rgb}{1.0,0.6,0}

\newcommand{\hM}{\color{dark-margenta}}

\usepackage{soul}

\newcommand{\dbr}[1]{\left\llbracket#1\right\rrbracket}

\newcommand{\BI}{{\bm B}_{I}}

\newcommand{\BF}{{\bm B}_{F}}

\newcommand{\Bp}{{\bm B}_{\mathrm{p}}}
\newcommand{\jI}{{\bm j}_{I}}

\newcommand{\jF}{{\bm j}_{F}}

\newcommand{\Tv}{\hat{\bm T}}
\newcommand{\Nv}{\hat{\bm N}}

\newcommand{\Mv}{\hat{\bm M}}
\newcommand{\et}{\hat{\bm \theta}}
\newcommand{\ep}{\hat{\bm \phi}}
\newcommand{\ero}{\hat{\bm \rho}}
\newcommand{\eo}{\hat{\bm \omega}}

\newcommand
{\Ov}
{\bm{\mathcal{O}}}
\newcommand
{\nv}
{\hat{\bm{n}}}

\newcommand{\Rv}{ \bm{\mathcal{R}} }

\newcommand{\Rvp}{ \bm{\mathcal{R}}^{\prime} }
\newcommand{\Rvpp}{ \bm{\mathcal{R}}^{\prime\prime} }

\newcommand{\bcd}
               { \bm{\cdot} }
\newcommand{\Rvd}
          { \bm{\dot{\mathcal R}} }
\newcommand{\Rvdst}
          { \bm{\dot{\mathcal{R}^{*}}} }
\newcommand{\Rvdd}
           { \bm{\ddot{\mathcal R}} }

\newcommand{\xv}{ \bm{x} }
\newcommand{\rv}{ \bm{r} }

\newcommand{\dl}{ \mathrm{d}l }

\newcommand{\fnu}{ \bm{f}_{\!\nu} }
\newcommand{\wnu}{ \bm{w}_\nu }
\newcommand{\bfnuIF}{ \bm{f}_{\!\nu_{\,I\!\!F}} }

\newcommand{\fnuIF}{ f_{\!\nu_{\,I\!\!F}} }

\newcommand{\bfnup}{ \bm{f}_{\!\nu_{\,\mathrm{p}}} }

\newcommand
{\Bu}
{B_{\mathrm{u}}}

\newcommand
{\Iu}
{I_{\mathrm{u}}}

\newcommand{\fnup}{ f_{\!\nu_{\,\mathrm{p}}} }

\newcommand{\f}{ \bm{f} }

\newcommand{\pd}[2]{\frac{\partial #1}{\partial #2}}
\newcommand{\Jnu}{{\bf J}_{\!\nu}}
\newcommand{\JnuIF}{{\bf J}_{\!\nu_{\,I\!\!F}}}
\newcommand{\Jnup}{{\bf J}_{\!\nu_{\,\mathrm{p}}}}
\newcommand{\Cst}{\mathcal{\mathcal C}^*}

\newcommand{\RBSL}{{\footnotesize R}BS{\footnotesize L}}

\begin{document}
%

\title{Optimization of Magnetic Flux Ropes Modeled with the \RBSL\ method} 

\correspondingauthor{Viacheslav S. Titov}
\email{titovv@predsci.com}

\author[0000-0001-7053-4081]{V. S. Titov}
\affiliation{Predictive Science Inc., 9990 Mesa Rim Road, Suite 170, San Diego, CA 92121} 

\author[0000-0003-1759-4354]{C. Downs}
\affiliation{Predictive Science Inc., 9990 Mesa Rim Road, Suite 170, San Diego, CA 92121} 

\author[0000-0003-3843-3242]{T. T\"{o}r\"{o}k}
\affiliation{Predictive Science Inc., 9990 Mesa Rim Road, Suite 170, San Diego, CA 92121} 

\author[0000-0003-1662-3328]{ J. A. Linker}
\affiliation{Predictive Science Inc., 9990 Mesa Rim Road, Suite 170, San Diego, CA 92121} 

\author[0000-0002-2633-4290]{ R. M. Caplan}
\affiliation{Predictive Science Inc., 9990 Mesa Rim Road, Suite 170, San Diego, CA 92121} 

\author[0000-0001-9231-045X]{ R. Lionello}
\affiliation{Predictive Science Inc., 9990 Mesa Rim Road, Suite 170, San Diego, CA 92121} 

\begin{abstract}
The so-called regularized Biot--Savart laws (\RBSL{s}) provide an efficient and flexible method for modeling pre-eruptive magnetic configurations of coronal mass ejections (CMEs) whose characteristics are constrained by observational images and magnetic-field data. This method allows one to calculate the field of magnetic flux ropes (MFRs) with small circular cross-sections and an arbitrary axis shape. The field of the whole configuration is constructed as a superposition of (1) such a flux-rope field and (2) an ambient potential field derived, for example, from an observed magnetogram. The \RBSL\ kernels are determined from the requirement that the MFR field for a straight cylinder must be exactly force-free.  For a curved MFR, however, the magnetic forces are generally unbalanced over the whole path of the MFR. To minimize these  forces, we apply a modified Gauss-Newton method to find  optimal MFR parameters. This is done by iteratively adjusting the MFR axis path and axial current. We then try to relax the resulting optimized configuration in a subsequent line-tied zero-beta magnetohydrodynamic simulation toward a force-free equilibrium. By considering two models of the sigmoidal pre-eruption configuration for the 2009 February 13 CME, we demonstrate how this approach works and what it is capable of. We show, in particular, that the building blocks of the core magnetic structure described by these models match to morphological features typically observed in such type of configurations. Our method will be useful for both the modeling of particular eruptive events and theoretical studies of idealized pre-eruptive MFR configurations.
\end{abstract}


\section{Introduction}
 \label{intro}

Coronal mass ejections (CMEs) are large eruptions of magnetized plasma from the Sun into the heliosphere, and the key agent of geomagnetic storms. It is very likely that most CMEs contain magnetic flux ropes (MFRs; e.g., \citealt{Chen2017}), at least for a substantial period of their lives. CMEs always originate in pre-eruptive configurations (PECs), which are closed magnetic fields low in the solar corona with a  current-carrying core region embedded in a largely potential field. Observations increasingly suggest that the core field is often comprised of, or contains, a nearly force-free MFR \citep[e.g.,][]{Canou2009, Green2009, Zhang2012, Patsourakos2013, Howard2014, Chintzoglou2015, Zhao2016, Wang2019}; see also the recent reviews by \cite{Liu2020} and \cite{Patsourakos2020}. This justifies the initialization of numerical models of CMEs with MFR configurations, both idealized  \citep[e.g.,][]{Amari2000, Fan2005, Aulanier2010, Torok2011} and constructed using observed magnetograms \citep[e.g.,][]{Manchester2008, Lugaz2011, Kliem2013, Amari2014, Inoue2018, Torok2018}; see also the review by \cite{Inoue2016}. 

CMEs result from the destabilization of the current-carrying substructures of PECs. The initial phase of CMEs can vary widely in terms of their acceleration \citep[e.g.,][]{Vrsnak2007}, rise direction \citep[e.g.,][]{Moestl2015}, or morphology \citep[e.g.,][]{Torok2010}. Furthermore, eruptions can be partial \citep[e.g.,][]{Gilbert2001} or even fully confined \citep[e.g.,][]{Ji2003}. Therefore, the accuracy of PEC reconstructions appears to be a crucial factor for a correct modeling of CMEs. An accurate reconstruction of PECs for observed cases, however, is particularly challenging \citep[][]{Patsourakos2020}, since the constraints required for the reconstruction can be inferred only indirectly from, e.g., observed filament shapes or the location of flare arcades or dimmings \citep[e.g.,][]{Palmerio2017}.

One way to produce observed equilibrium PECs is by trying to form MFRs via a slow boundary-driven evolution, which can be magnetofrictional \citep[e.g.,][]{Cheung2012, Price2020} or magnetohydrodynamic \citep[MHD; e.g.,][]{Lionello2002, Bisi2010, Zuccarello2012, Jiang2016, Hayashi2018}. This requires the design of photospheric boundary conditions that emulate the physical processes on the Sun leading to PECs, and at the same time are consistent with observed magnetograms and morphological features. This approach is non-trivial, even if one tries to match only the normal component of the photospheric magnetic field. It may, therefore, take many trial-and-error attempts to create a stable PEC that satisfactorily matches the observations---especially because this method has no simple means to control the detailed properties and stability of the resulting MFR.

Another way to reconstruct PECs is via non-linear force-free field (NLFFF) extrapolations \citep[e.g.,][]{Schrijver2008}, by using vector magnetic data. However, these data are measured at the photospheric level, where the characteristic gradients of plasma pressure are not negligible compared to the magnetic forces, corresponding to a significant deviation of the field from a force-free state.  Therefore, the data must be ``pre-processed'' to be compatible with the extrapolation higher up in the chromosphere and low corona \citep{Wiegelmann2006}. Vector magnetic field observations also suffer from noise and disambiguation issues, which makes the reconstruction of PECs based on NLFFF extrapolations a non-trivial problem as well.

An alternative to these boundary-condition-based methods is the MFR insertion method \citep{vanBall2004, Su2011, Savcheva2012}, which uses observations of filaments, loops, etc., to directly constrain the field model. In this method, PEC equilibria are constructed via a two-step iterative procedure. First, following the desired MFR shape inferred from the observations, a field-free cavity is  constructed within the background potential magnetic field. The cavity is then filled with axial and azimuthal magnetic flux, such that the corresponding electric current is fully neutralized. Second, the resulting MFR configuration is subjected to a magnetofrictional relaxation toward a force-free equilibrium, which may contain net currents. These two steps are repeated by varying the inserted magnetic fluxes and/or the cavity shape, until a suitable equilibrium is reached. Note that all configurations obtained after step one are, by construction, far from an equilibrium state. Therefore, the relaxation can significantly change the inserted MFR and its adjacent magnetic structure. This makes the properties of the modeled PECs difficult to control, and many iterations of trial and error may be required to reach the desired result.

Fortunately, this situation can be substantially improved by using regularized Biot--Savart laws (\RBSL{s}), which we recently proposed for constructing PECs with embedded MFRs \citep[][]{Titov2018}. Similarly to the MFR insertion method, our \RBSL\ method uses observations to constrain the shape of the modeled MFR. Additionally, however, it employs the ambient potential field at the apex point of the MFR axis to estimate the axial current. The estimation is obtained from the condition that the components of the ambient field and the field generated by the axial current cancel each other at this point in the plane perpendicular to the axis. For thin MFRs, the field produced by the azimuthal current is small at the apex point and, therefore, neglected. Except for this simplification, the estimation is done in the same way as for the Titov \& D\'emoulin-modified (TDm) MFR model described in \cite{Titov2014}. 

In both approaches, the whole configuration is represented by a linear superposition of the MFR and ambient potential fields. However, while the TDm model describes the field for MFRs of arc shape only, the \RBSL{s} do so for MFRs of arbitrary shape. Therefore, by using the \RBSL{s} one can model a wider class of PECs with a more complex geometry of embedded MFRs.

For such cases, however, the estimation of the axial current via the ambient field at the axis apex may not be accurate enough. This is because, in general, the ambient and MFR fields vary along the axis path. It is desirable, hence, to adjust the MFR shape in such a way that the equilibrium condition holds not only at the apex point, but also at all other points along the axis path.

For this purpose, we propose here an efficient procedure that extends the equilibrium condition to the whole path of the MFR and additionally generalizes it  in the following two ways. First, we incorporate the previously neglected influence of the azimuthal current on the MFR equilibrium. Second, we take into account the variations of the current density and magnetic field across the MFR, which were also discarded before. Both generalizations become important at path segments where the local curvature radius is comparable to the radius of the MFR.

This procedure allows one to determine the unbalanced magnetic forces at several cross-sections of the MFR, and to minimize them by optimizing the shape of the axis path and the axial current of the MFR, using the Least Squares method. The residual magnetic stress in the optimized PEC can then be relaxed via  magnetohydrodynamic (MHD) simulations performed under vanishing plasma pressure ($\beta=0$) and photospheric line-tying conditions. 

Note that the optimization adjusts not just the axis path, but also the connectivity of magnetic field lines to the photospheric boundary. This can be important when seeking agreement with observed morphological features of the PEC, primarily because the initial guess for the MFR axis may not be perfect. Furthermore, the adjusted magnetic connectivity and minimized magnetic stress control, to a large extent, the subsequent line-tied MHD relaxation and thus the properties of the resulting PEC. In this way, the optimization allows one to reach the desired result with fewer trial-and-error attempts and less sensitivity to initial choices. 

In addition to the optimization procedure, we present an improved formulation of the \RBSL{s}, which reduces the normal component of the MFR field at the photosphere (including the footprint areas) to negligibly small values. This provides a much closer match of observed magnetograms than our previous formulation, which is essential for event case-studies. Ultimately, we believe this combination of optimization, relaxation, and normal-matching using \RBSL{s}, can greatly facilitate the rapid construction of realistic, stable, and highly energized PECs. We expect that our new method will also be useful for a variety of non-CME applications, such as the recently developed MHD models
of prominence formation \citep[e.g.,][]{Xia2014b, Fan2019}.

In Section \ref{s:M}, we fully describe our method, including the improved \RBSL{s} (Section \ref{s:AIAF}) and the basic theory used for the proposed optimization procedure (Sections \ref{s:path}-\ref{s:opt}). We then illustrate in Section \ref{s:examples} the method by applying it to the modeling of PECs for the 2009 February 13 CME. Section \ref{sum} summarizes our results, and the Appendix provides an auxiliary mathematical framework for calculating magnetic fields and current densities defined by the \RBSL{s}.


\section{Method
\label{s:M}}

\subsection{Improved \RBSL{s}
\label{s:AIAF}
}

In this work, we model any PEC prior to its relaxation as the following superposition of three different components of the magnetic field:
\begin{eqnarray}
\bm{B}
&=&
\Bp
+
\BI
+
\BF
\, , 
	\label{B}
\end{eqnarray}
where $\Bp$ represents a potential magnetic field corresponding to a certain distribution of the radial magnetic field, which is derived, for example, from observations. The other two components, $\BI$ and $\BF$, are, respectively, azimuthal and axial magnetic fields generated by axial net current $I$ and axial net flux $F$ of a thin MFR. These components, in turn, are
\begin{eqnarray}
\BI
&=&
\nabla
\times
\bm{A}_I
	\label{BI}
\, ,
\\
\BF
&=&
\nabla
\times
\bm{A}_F
	\label{BF}
\, ,
\end{eqnarray}
where $\bm{A}_I$ and $\bm{A}_F$ are axial and azimuthal vector potentials, respectively, defined relative to the axis of MFR. The latter is a closed curve formed by coronal and subphotospheric paths, $\mathcal{C}$ and $\Cst$, respectively, and represented by a vector  $\Rv(l)$ that depends on the arc length $l$ of the curve. For these vector potentials, we adopt here the \RBSL{s} proposed earlier in \citep{Titov2018} by assuming that our MFR has a constant cross-sectional radius $a$. Then the axial vector potential at a given point $\xv$ is described by 
\begin{eqnarray}
  {\bm A}_I(\xv)
  &=&
  \int_{\mathcal{C}\, \cup\,\Cst}
  K_I(r)
  \;
  \Rvp
  \;
  \frac{\dl}{a}
  \, , 
\quad
\dbr{
\frac
{\mu I}
{4\pi}
}
  \, , 
	\label{AI}
\end{eqnarray}
where $\rv \equiv \rv(l) = \left( \xv -\Rv(l) \right) / a $, $\Rvp ={\mathrm d} \Rv/{\mathrm d}l$ is a unit vector tangential to the axis path. The double brackets henceforth contain the unit in which the value displayed on the left from it is measured. The \RBSL\ kernel of ${\bm A}_I$ is
\begin{eqnarray}
K_{I}(r)
=
\begin{cases}
\frac{2}{\pi} 
\left(
\frac
{\arcsin r}
{r}
+
\frac
{
5-2\, r^2
}
{
3
}
\sqrt{1-r^2}
\right)
\, ,
& \text{$r \in (0,1]$,}
\\
r^{-1}
\, ,
& \text{$r>1$,}
\end{cases}
  \label{KI}
\end{eqnarray}
whose domain of definition smoothly extends to $r=0$ with
\begin{eqnarray}
\lim
_{
r
\rightarrow
0^{+}
}
K_{I}(r)
=
16
/
(3\,\pi)
\approx 
1.698
\, .
\end{eqnarray}
In the limit of vanishing curvature of the MFR, $K_{I}(r)$ by construction provides the azimuthal magnetic field given by
\begin{eqnarray}
B_{\mathrm{az}}(\rho)
&=&
2
\rho 
\left(
2
-\rho^2
\right)
\, ,
\quad
\rho
\in
[0,1]
\, ,
\qquad
\dbr
{
\frac
{\mu I}
{4 \pi a}
}
\, ,
   \label{Baz}
\end{eqnarray}
where $\rho$ is the distance measured from the axis of the MFR and normalized to $a$. This field corresponds to a parabolic profile of the axial current density, namely,
\begin{eqnarray}
j_\mathrm{ax}(\rho) 
&=&
2
\left(
1
-
\rho^2
\right)
\, ,
\quad
\rho
\in
[0,1]
\, ,
\qquad
\dbr
{
\frac
{I}
{\pi a^2}
}
\, .
    \label{jax}
\end{eqnarray}
We assume in this paper that the closure of  the coronal current $I$ flowing along the path $\mathcal{C}$ is reached via a fictitious subphotospheric path $\Cst$ that is simply a mirror image of $\mathcal{C}$ about the solar surface (see Section \ref{s:path}). This constraint on the shape of the path allows one to nearly vanish the resulting normal component of  $\BI$ at the solar surface.

To cause a similar effect to the $\BF$--distribution at the boundary, we define the azimuthal vector potential as
\begin{eqnarray}
{\bm A}_F(\xv)
&=&
\int_{\mathcal{C}}
K_F(r)
\,
\Rvp
\times
\rv
\;
\frac{\dl}{a}
\nonumber
\\
&-&
\int_{\Cst}
K_F(r)
\,
\Rvp
\times
\rv
\;
\frac{\dl}{a}
  \, , 
\quad
\dbr
{
\frac
{
F
}
{
4
\pi
a
}
}
  \, .
	\label{AF}
\end{eqnarray}
This expression implies that the corresponding axial fluxes flow along $\mathcal{C}$ and $\Cst$ in opposite directions. Since $\Cst$ is a mirror image of $\mathcal{C}$ about the solar boundary, these fluxes meet at the same angles to the boundary and thereby cancel each other out. Due to this trick, the resulting normal component of $\BF$ also becomes negligible at the solar surface.

The \RBSL\ kernel of ${\bm A}_F$ is
\begin{eqnarray}
K_{F}(r)
=
\begin{cases}
\frac
{2}
{
\pi
r^2
}
\left(
\frac
{\arcsin r}
{r}
-
\sqrt{1-r^2}
\right) 
+
\frac
{2}
{\pi}
\sqrt{1-r^2}
\nonumber
\\
+
\frac
{
5
-
2\,
r^2
}
{2
\sqrt{6}
}
\left[
1
-
\frac
{2}
{\pi}
\arcsin
\left(
\frac
{
1
+
2\,
r^2
}
{
5
-
2\,
r^2
}
\right)
\right]
\, ,
& \text{$r \in (0,1]$,}
\\
r^{-3}
\, ,
& \text{$r>1$,}
\end{cases}
\\
	\label{KF}
\end{eqnarray}
whose domain of definition smoothly extends to $r=0$ with
\begin{eqnarray}
\lim
_{
r
\rightarrow
0^{+}
}
K_{F}(r)
&=&
\frac
{10}
{3\pi}
+
\frac
{5}
{
2
\sqrt{6}
\pi
}
\left[
\pi
-
2\,
\arcsin
\left(
\frac
{1}
{5}
\right)
\right]
\nonumber
\\
&\approx&
1.951
\, .
	\label{KF(0)}
\end{eqnarray}
In the limit of vanishing curvature of the MFR, $K_{F}(r)$ by construction provides the axial magnetic field and and azimuthal current density given, respectively, by
\begin{eqnarray}
B_\mathrm{ax}(\rho)
&=&
\frac
{20}
{
3
\sqrt{3}
}
\left(
1
-
\rho^2
\right)
\sqrt{
5
-
2
\,
\rho^2
}
\, ,
\quad
\rho
\in
[0,1]
\, ,
\quad
\dbr
{
\frac
{F}
{
4
\pi
\,
a^2
}
}
\, ,
\nonumber
\\
   \label{Bax}
\\
j_\mathrm{az}(\rho)
&=&
\frac
{
40
\,
\rho
}
{
\sqrt{3}
}
\frac
{
\left(
2
-
\rho^2
\right)
}
{
\sqrt
{
5
-
2
\,
\rho^2
}
}
\, ,
\quad
\rho
\in
[0,1]
\, ,
\quad
\dbr
{
\frac
{F}
{
4
\pi
\mu
\,
a^3
}
}
\, .
	\label{jaz} 
\end{eqnarray}
It is easy to check that a straight cylindrical MFR  defined by Eqs. (\ref{Baz}), (\ref{jax}), (\ref{Bax}), and (\ref{jaz}) is indeed force-free if the net axial flux and current are related by
\begin{eqnarray}
\frac
{F}
{
\mu
I
a
}
\equiv
\sigma
=
\frac
{\pm 3}
{5\sqrt{2}}
\approx
0.424
\, ,
	\label{Fpb}
\end{eqnarray}
where the positive and negative signs correspond to right- and left-handed twist (chirality) of the MFR, respectively.

We assume that this relationship also holds true for a curved MFR described by Eqs. (\ref{AI}), (\ref{KI}), (\ref{AF}), and (\ref{KF}). In this way, we manage to keep the resulting configuration as close as possible to an equilibrium for sufficiently thin MFRs, which quantitatively means that
\begin{equation}
\kappa
a
\ll
1
	\label{ka}
\end{equation}
along the axis path of curvature $\kappa(l)$. The appropriate power-laws decay of $K_{I}(r)$ and $K_{F}(r)$ at $r>1$ ensures that, externally, our MFR manifests itself as a current and flux carrying thread described by classical Biot--Savart laws \citep{Jackson1962}.

\subsection{Axis Path Model of the MFR
	\label{s:path}}

A special scrutiny is required for constructing a discretized model of the axis path $\mathcal{C}$ to make its optimization process stable. We represent $\mathcal{C}$ in terms of a cubic spline that smoothly join $N+1$ points, $\Rv_0,\ldots,\Rv_N$, called control nodes. Instead of $l$, it is convenient to parameterize $\mathcal{C}$ by a continuous parameter $\nu$ whose values coincide with the numbers  $0,\ldots,N$ at the control nodes. Any other point $\Rv(\nu)$ of $\mathcal{C}$ is determined then by the vector function
\begin{eqnarray}
\Rv(\nu)
&=&
\sum^{N}_{i=0}
S^{N}_i(\nu)
\,
\Rv_i
\, ,
\quad
\nu
\in
[0,N]
\, ,
	\label{Rvnu}
\end{eqnarray}
in which $S^{N}_i(\nu)$ is a piecewise cubic polynomial of $\nu$ associated with the i-th control node, namely,
\begin{eqnarray}
S^{N}_i(\nu)
&=&
\begin{cases}
\sum
\limits_{m=0}^{3}
c^{1}_{i,m}
\nu^{m}
,
\quad
0
\le
\nu
<
1,
\\
\qquad
\ldots
\\ 
\sum
\limits_{m=0}^{3}
c^{j}_{i,m}
\nu^{m}
,
\quad
j-1
\le
\nu
<
j,
\\
\qquad
\ldots
\\ 
\sum
\limits_{m=0}^{3}
c^{N}_{i,m}
\nu^{m}
,
\quad
N-1
\le
\nu
\le
N\,.
\end{cases}
	\label{SNi}
\end{eqnarray}
Its coefficients $c^{j}_{i,m}$, $j=1,\ldots,N$, are uniquely defined by a linear system of $4N$ equations.
This system consists of the following $2N$ equations:
\begin{eqnarray}
S^{N}_i(j-1)
&\equiv&
\sum
\limits_{m=0}^{3}
(j-1)^{m}
c^{j}_{i,m}
=
\begin{cases}
1 & \text{if $i=j-1$,}\\
0 & \text{otherwise,}
\end{cases}
	\label{ce1}
\\
S^{N}_i(j)
&\equiv&
\sum
\limits_{m=0}^{3}
j^{m}
c^{j}_{i,m}
=
\begin{cases}
1 & \text{if $i=j$,}\\
0 & \text{otherwise,}
\end{cases}
	\label{ce2}
\end{eqnarray}
supplemented with $2(N-1)$ continuity conditions at $\nu = j = 1,\ldots,N-1$ for the first and second 
derivatives of $S^{N}_i(\nu)$,
\begin{eqnarray}
\sum
\limits_{m=1}^{3}
m
j^{m-1}
\left(
c^{j}_{i,m}
-
c^{j+1}_{i,m}
\right)
=
0
\,
,
	\label{ce3}
\\
c^{j}_{i,2}
-
c^{j+1}_{i,2}
+
3
j
\left(
c^{j}_{i,3}
-
c^{j+1}_{i,3}
\right)
=
0
\,
,
	\label{ce4}
\end{eqnarray}
respectively, and  two simple or natural endpoint conditions of vanishing second derivatives of $S^{N}_i(\nu)$ at $\nu=0$ and $\nu=N$:
\begin{eqnarray}
c^{1}_{i,2}
=
0
\,
,
	\label{ce5}
\\
c^{N}_{i,2}
+
3N
c^{N}_{i,3}
=
0
\,
.
	\label{ce6}
\end{eqnarray}

\begin{figure}[ht!]
\plotone{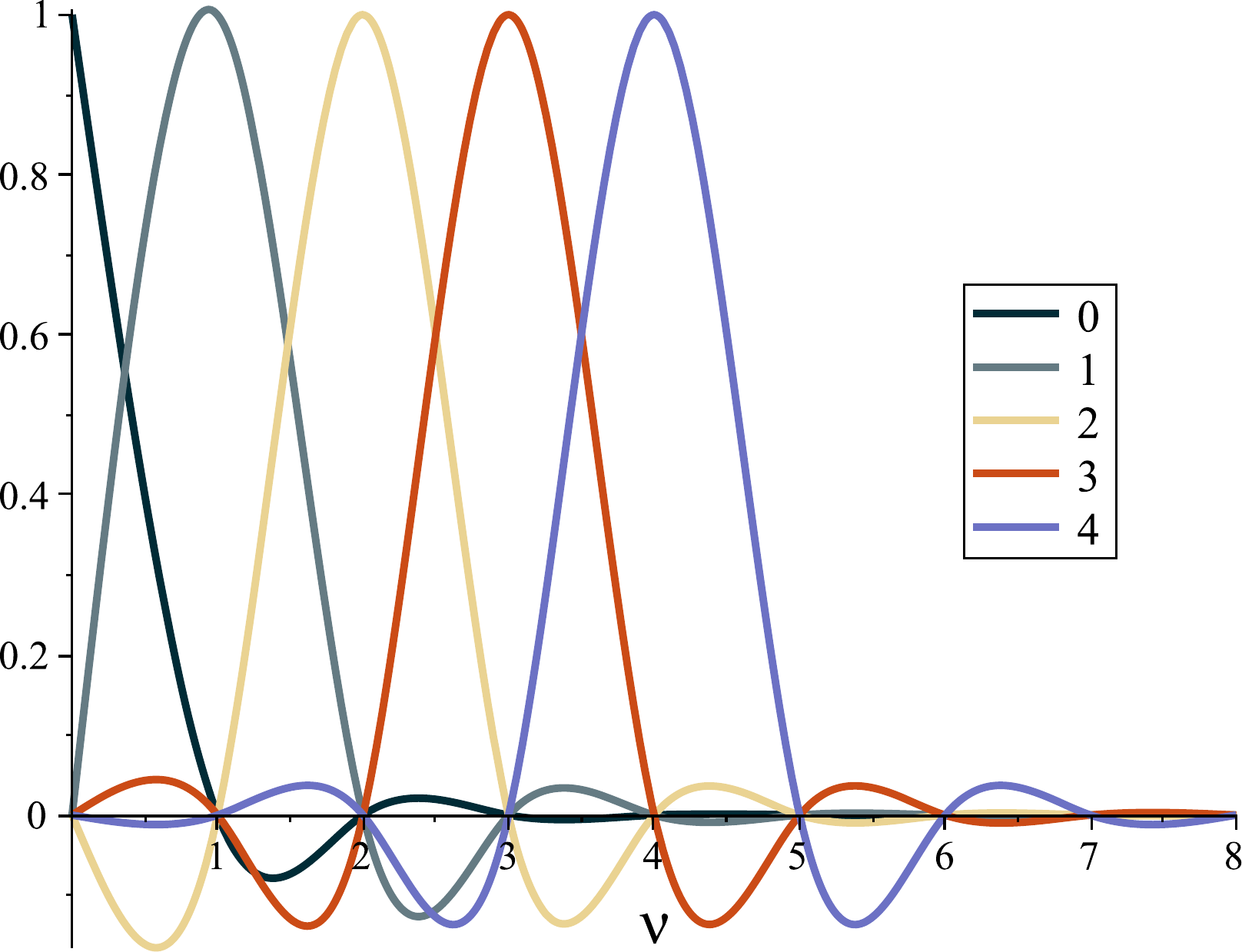}
\caption{
Basic cubic spline function $S^{N}_i(\nu)$ associated with the $i$-th control node for $N=8$ and $i=0,\ldots,4$.
	\label{f:SNi(nu)}}
\end{figure}

The defined spline $S^{N}_i(\nu)$ describes the contribution of the $i$-th node  to the shape of the axis path under unchanged positions of the other $N$ nodes (see Figure \ref{f:SNi(nu)}).
As Eq. (\ref{Rvnu}) explicitly states, this contribution is linearly proportional to the vector $\Rv_i$ representing the $i$-th control node.
It is smoothly distributed over the axis path with a maximum at this node and decays relatively fast by having smaller and smaller maximums between other nodes.
This representation of the axis-path spline allows one to assess the influence of the position of a single node on the whole shape of the path.

The arc length $l(\nu)$ of  $\mathcal{C}$ as a function of $\nu$ is a solution of the following ordinary differential equation (ODE):
\begin{eqnarray}
\frac
{
\mathrm{d}
l 
}
{
\mathrm{d}
\nu
}
&=&
{\bigl|\Rvd(\nu)\bigr|}
\, ,
	\label{nu2l}
\end{eqnarray}
in which
\begin{eqnarray}
\Rvd(\nu)
&=&
\sum^{N}_{i=0}
\frac
{
\mathrm{d}
\ 
}
{
\mathrm{d}
\nu
}
S^{N}_i(\nu)
\,
\Rv_i
	\label{dRvnu}
\end{eqnarray}
is the $\nu$--derivative of   Eq. (\ref{Rvnu}). By integrating Eq. (\ref{nu2l}) from $0$ to $N$, one obtains the total arc length $L$ of $\mathcal{C}$.
The inverted Eq. (\ref{nu2l}) yields another ODE
\begin{eqnarray}
\frac
{
\mathrm{d}
\nu 
}
{
\mathrm{d}
l
}
&=&
\frac
{1}
{\bigl|\Rvd(\nu)\bigr|}
	\label{l2nu}
\end{eqnarray}
whose solutions define the inverse relationship $\nu(l)$ between $l$-- and $\nu$--parameterizations of the curve ${\mathcal C}$.
%
\begin{figure}[ht!]
\plotone{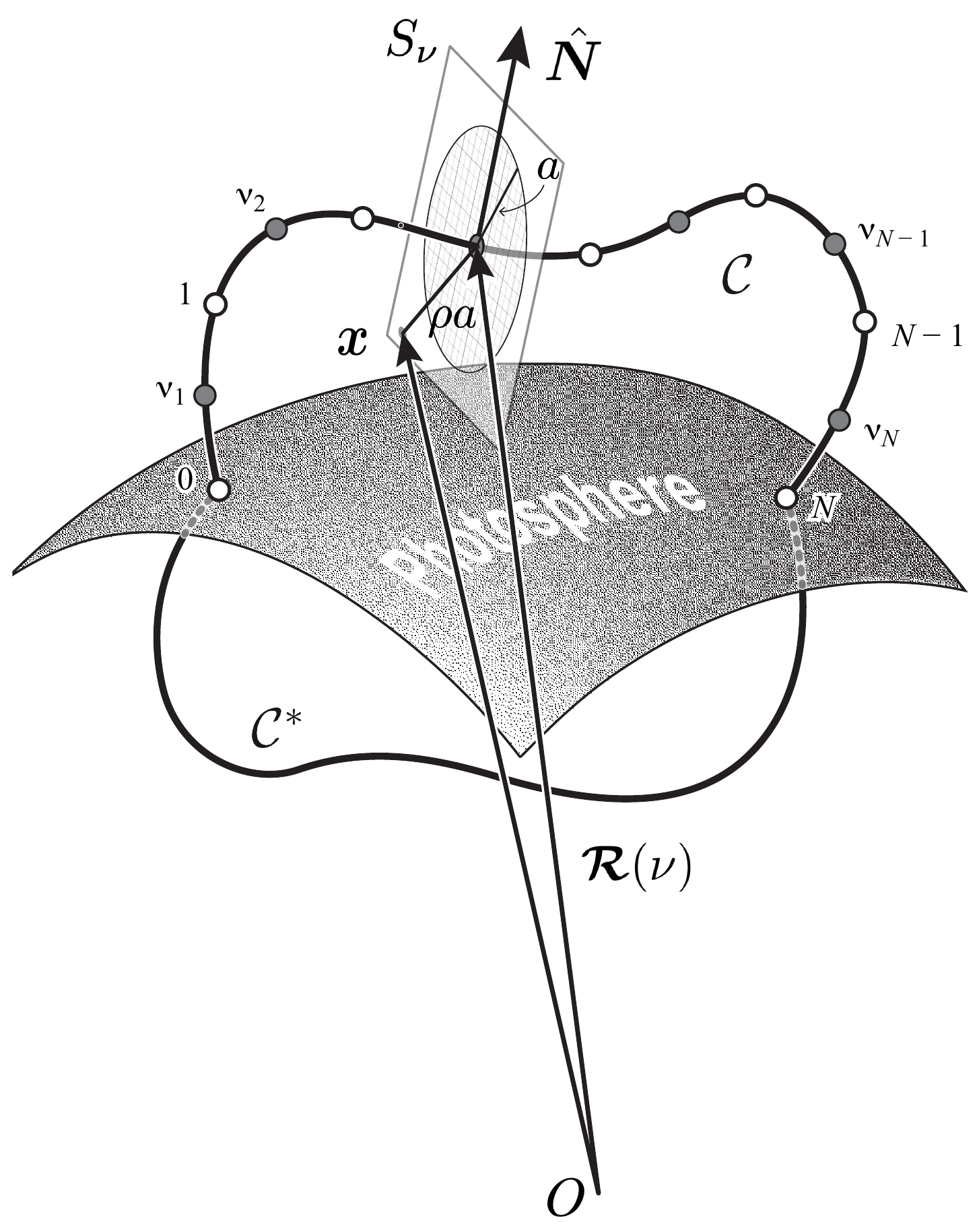}
\caption{The coronal  axis path ${\mathcal C}$ is represented by a vector function $\Rv(\nu)$, which is defined in terms of a cubic spline of $N+1$ equidistant control nodes $\Rv_i$ (white circles) uniformly parameterized by parameter $\nu$ from $0$ to $N$.
The gray circles show evaluation nodes at which the line density $\fnu$ of the magnetic force is calculated by Eq. (\ref{f*}) via the integral of the Lorentz force, taken with a certain weight, over the corresponding cross-sections $S_\nu$ perpendicular to the path. The subphotospheric axis path ${\mathcal C}^{*}$ is a copy of ${\mathcal C}$ mirrored about a plane that locally approximates the spherical solar boundary.
	\label{f:cross-section}}
\end{figure}

The described ODEs help us to keep the control nodes $\Rv_i$ equidistant along the path at each iteration toward its optimized shape, which brings some kind of stiffness to the  path during its deformation. Operationally, we determine first $L$, as described above, and then integrate Eq. (\ref{l2nu}) from $0$ to $l_j\equiv j\, L/N$ to obtain $\nu_j = \nu(l_j),\ j=1,\ldots,N-1$. To prevent a deterioration of the path model due to an excessive separation of the nodes, we reset $N$ at each iteration by the rounded $\max(N,L/\Delta l_\mathrm{max})$, where $\Delta l_\mathrm{max}\sim a$ is a maximally allowable arc length between the nodes.

The evaluation of Eq. (\ref{Rvnu}) at  $\nu = \nu_j$ yields equidistant path points $\Rv_j$ that are used as new control nodes in the same Eq. (\ref{Rvnu}). The newly defined path is slightly different than the starting one, so their arc lengths between the control nodes are different as well. Nevertheless, the $\Rv_j$-nodes at the new path appear for the used $N$ more equidistant than the $\Rv_i$-nodes at the starting path. Repeating such a procedure of resampling control nodes, one can make them  more and more equidistant at the modeled path $\Rv(\nu)$. In fact, the application of this procedure shows that, after each sufficiently small deformation of the path, one such resampling might be enough to make the optimization procedure  stable (Section \ref{s:opt}). The axis path defined by Eq. (\ref{Rvnu}) with equidistant control nodes is henceforth called {\it canonical}.

The control nodes, different from the foot points of the path,  are allowed to be displaced only along the binormal $\Mv$ and normal  $\Nv$ of the Frenet-Serret basis of the path. For our path model, they are calculated by using Eq. (\ref{dRvnu}) and
\begin{eqnarray}
\Rvdd(\nu)
&=&
\sum^{N}_{i=0}
\frac
{
\mathrm{d^2}
\ 
}
{
\mathrm{d}
\nu^2
}
S^{N}_i(\nu)
\,
\Rv_i
	\label{d2Rvnu}
\end{eqnarray}
as follows:
\begin{eqnarray}
\Tv
&=&
\Rvp
\equiv
\frac
{
\mathrm{d}
\Rv
}
{
\mathrm{d}l
}
=
\frac
{
\Rvd
}
{
\left|
\Rvd
\right|
}
\, ,
	\label{Tv}
\\
\Nv
&=&
\frac
{
\Rvpp
}
{
\left|
\Rvpp
\right|
}
\, ,
\qquad
\kappa
\equiv
\left|
\Rvpp
\right|
\, ,
	\label{Nv}
\\
\Rvpp
&\equiv&
\frac
{
\mathrm{d}^2
\Rv
}
{
\mathrm{d}l^2
}
=
\frac
{
\Rvdd
}
{
\Rvd
^2
} 
-
\frac
{
\Rvd
\bcd
\Rvdd
}
{
\Rvd
^4
}
\,
\Rvd
\, ,
	\label{Rvpp}
\\
\Mv
&=&
\Tv
\times
\Nv
\, .
	\label{Mv}
\end{eqnarray}
By using these expressions evaluated at $\nu=\nu_j$, one can obtain the normal $\Nv_j$ and binormal $\Mv_j$ along which the corresponding control nodes $\Rv_j,\ j=1,\ldots,N-1,$ are displaced in the optimization process.

In contrast, the control nodes $\Rv_0$ and $\Rv_N$ as being the foot points of the path are allowed to be displaced strictly along the solar surface. This implies that
\begin{eqnarray*}
&&
\Mv_0 = \et_{\mathrm{FP}1}, \quad
\Mv_N = \et_{\mathrm{FP}2}
\, , 
\\
&&
\Nv_0 = \ep_{\mathrm{FP}1}, \quad
\Nv_N = \ep_{\mathrm{FP}2}
\, ,
\end{eqnarray*}
where $\et$ and $\ep$ with the corresponding subscripts are the unit coordinate vectors at the foot points of the path defined in spherical coordinate system $(r,\theta,\phi)$ with the origin at the center of the Sun. Thus, a small variation of the axis path described by Eq. (\ref{Rvnu}) can be written as 
\begin{eqnarray}
\delta\!\Rv
&=&
a
\sum^{N}_{i=0}
S^{N}_i(\nu)
\,
\left(
\xi_i
\Mv_i
+
\eta_i
\Nv_i
\right)
\, ,
	\label{dR}
\end{eqnarray}
where the node displacements $\xi_i$ and $\eta_i $ are normalized to $a$. We hold these displacements to be $\ll 1$ through all iterations of the optimization process.

The subphotospheric axis path ${\mathcal C}^{*}$ at each iteration is chosen to be a copy of ${\mathcal C}$ mirrored about the plane that passes through the foot points $\Rv_0$ and $\Rv_N$ and has the normal
\begin{equation}
\nv
=
\Ov
/
\left|
\Ov
\right|
\, ,
	\label{nv}
\end{equation}
where
\begin{equation}
\Ov
=
(\Rv_0 + \Rv_N)/2
	\label{Ov}
\end{equation}
is an average of the foot points.
For $\left| \Rv_N - \Rv_0 \right| \ll R_\sun$, this plane well approximates a  plane touching the solar surface at the point $R_\sun \nv$.
The corresponding mirror images of the control nodes are given by
\begin{eqnarray}
\Rv_i^{*}
=
\Rv_i
-
2
\,
\nv
\bcd
\left(
\Rv_i
-
\Ov
\right)
\nv
\, .
	\label{Rvist}
\end{eqnarray}
By using the same direction and type of parameterization as for the path $\mathcal{C}$ (Eq. (\ref{Rvnu})), one can determine its corresponding mirrored points at $\Cst$ from
\begin{eqnarray}
\Rv^{*}(\nu)
&=&
\sum^{0}_{i=N}
S^{N}_i(\nu)
\,
\Rv_i^{*}
\, ,
\quad
\nu
\in
[0,N]
\, .
	\label{Rvnust}
\end{eqnarray}
This closure of $\mathcal{C}$ makes it  possible to minimize the normal component of magnetic field that  the flux-rope currents produce at the boundary, as discussed in Section \ref{s:AIAF}. The normal component of the resulting field    then is almost due to the potential field $\Bp$ (see Eq. (\ref{B})), or, in other words, it becomes almost identical to the component derived from observations. A difference between them is only due to the curvature of the solar surface, which is relatively small for typical source regions of CMEs.
The configurations of larger size require a more sophisticated approach, which we will consider in a further publication.

\subsection{Line Density of the Residual Magnetic Force
	\label{s:fnu}}

In order to estimate how far our approximate MFR PEC deviates from an equilibrium, we have to determine the line density of the residual magnetic forces along MFR, or, in other words, the magnetic force $\fnu$ per unit length of the MFR.
Its expression can rigorously be derived by using {\hM the} Maxwell stress tensor integrated over the surface of an elementary wedge of the rope.
The latter is formed by slicing the MFR with two planes that are perpendicular to the axis and separated from each other along the axis by a segment of length $\Delta l$ that is tending to zero.
The lateral surface of the wedge has to fully enclose the part of the MFR sliced by these planes.
The corresponding integral over the boundary of the wedge divided by $\Delta l$ provides{\hM,} after some lengthy algebra, the required expression of $\fnu$.
The form of this expression suggests, however, that $\fnu$ can be obtained in a much easier manner by simply integrating the volumetric density of the Lorentz force $\bm{j\times B}$ over the volume of the elementary wedge.
Let $\mathrm{d}^2 \xv$ be a surface element at a point $\xv$ on one of the wedge planes containing a cross-section $S_\nu$ of the MFR  (see Figure \ref{f:cross-section}).
The wedge width is $(1-\kappa Y)\, \Delta l$, where $Y$ is the coordinate value measured along the normal $\Nv$ from the point of intersection of $S_\nu$ with the axis path.
Then one obtains from Eq. (\ref{Nv}) that this coordinate is $Y=(\xv - \Rv) \bcd \Rvpp$,
so that the corresponding volume element of the wedge is $\left[1 - (\xv - \Rv) \bcd \Rvpp \right]\, \Delta l\, \mathrm{d}^2 \xv$.
Thus, the required expression for the line density of the Lorentz force is
\begin{eqnarray}
\fnu
=
\int_{S_\nu}
\left[
1
-
(\xv - \Rv)
\bcd
\Rvpp
\right]
\,
(\bm{j\times B})
\:
\mathrm{d}^2
\xv
\, ,
	\label{f}
\end{eqnarray}
where $\Rv$ and $\Rvpp$ are given for our path model by Eqs. (\ref{Rvnu}) and (\ref{Rvpp}).

This consideration also allows one to see  that the concept of $\fnu$ itself physically correct only for the axis points, where
$\kappa
a
\ll
1
$
holds true, or, in other words, the MFR is locally thin. If $\kappa a$ becomes larger than $1$ at some point (i.e., the MFR is locally thick), the corresponding center of curvature of the path turns out to be inside the rope, so that the corresponding cross-sectional planes start slicing the MFR into two elementary wedges. To cover, at least formally, such ``corner'' points, Eq. (\ref{f}) should be extended via the following modification:
\begin{eqnarray}
\fnu
=
\int_{S_\nu}
\left|
1
-
(\xv - \Rv)
\bcd
\Rvpp
\right|
\,
(\bm{j\times B})
\:
\mathrm{d}^2
\xv
\, ,
	\label{f*}
\end{eqnarray}
where we apply the modulus to the metric factor to handle both wedges on equal footing, as their contributions to $\fnu$ are similar.

Although this extension generally covers such ``corner'' points, it is desirable to prevent the formation of these points in the optimization process of the path for other reason. Note that, for our \RBSL\ flux rope, the normalized current density is
\begin{eqnarray}
\bm{j}
=
\jI
+
\sigma
\jF
\, ,
\qquad
\dbr
{
\frac
{I}
{4\pi a^2}
}
\, ,
	\label{jv}
\end{eqnarray}
where $\jI$ and $\jF$ are normalized axial and azimuthal current densities described by  Eqs. (\ref{jI})--(\ref{KjI2}) and (\ref{jF_int})--(\ref{KjF}), respectively. Appendix \ref{s:JFtorus} demonstrates that, for $\kappa a \rightarrow 1$, a singularity is developed in $\jF$--distribution  at the concave side of MFR, which signifies the condition of being avoided when applying the \RBSL\ method.

Such a sensitivity of the method to  $\kappa a \sim 1$  motivated us to evaluate $\fnu$ for the purpose of optimization  at points different from the control nodes $\Rv_i$, because $\kappa a$ tends to have local maxima at $\Rv_i$. For the evaluation of $\fnu$, therefore, we choose the points that are equidistantly separated from the nearest control nodes. To determine them, we first calculate the corresponding values  $\nu_i$, $i=1,\ldots,N$, by using Eqs. (\ref{nu2l})--(\ref{l2nu}) and then evaluate Eq. (\ref{Rvnu}) at $\nu=\nu_i$ to obtain  the required evaluation nodes, i.e, in a similar way as the control nodes are obtained before (see Section \ref{s:path}).

Let the potential field $\Bp$ and axial current $I$ be measured in $\Bu$ and $\Iu$ units, respectively, such that
\begin{eqnarray}
\Iu
=
4\pi a
\Bu/\mu
\end{eqnarray}
and so
\begin{eqnarray}
I
=
C_I
\Iu
\, ,
\end{eqnarray}
where the dimensionless coefficient $C_I$ is yet to be determined in further optimization.
Then $\fnu$ can be written as
\begin{eqnarray}
\fnu
&=&
C_I
\bfnup
+
C_I^2
\bfnuIF
\, ,
\qquad
\dbr
{
\frac
{
\Bu^2
}
{\mu}
a
}
\, ,
	\label{fnunorm}
\end{eqnarray}
where $\bfnup$ and $\bfnuIF$ are two separate parts of $\fnu$ due to $\bm{j \times}\Bp$  and $\bm{j \times}(\BI + \sigma \BF)$, respectively. The current density and magnetic field components are calculated here by using differential \RBSL\ formulations described in Appendix \ref{s:diff_RBSL-forms}. The latter allows one to represent our \RBSL\ integrals as solutions of certain ODEs, which in turn makes it possible to exploit the power of the adaptive step refinement in the existing ODE solvers.

\subsection{Optimization of the MFR Parameters
	\label{s:opt}}

We constructed several metrics for measuring how far from equilibrium an MFR configuration is, and used them as cost functions in a minimization procedure to obtain approximate equilibria. The construction invokes the nonlinear least squares method, and the corresponding minimization is performed iteratively by varying independent MFR parameters, namely the axial current and the coordinates of the control nodes.
More precisely, we vary the dimensionless parameter $C_I$ and the $2(N+1)$--dimensional vector of the node displacements
\begin{eqnarray}
\bm{\chi}
&=&
\left(
\xi_0
,
\ldots
,
\xi_{N}
,
\eta_0
,
\ldots
,
\eta_{N}
\right)^{\mathrm T}
\, ,
	\label{chi}
\end{eqnarray}
where the subscript $\mathrm{T}$ denotes matrix transposition.
Note that the axial flux $F$ in our \RBSL\ approach is not an independent parameter; it scales with the axial current $I$ according to Eq. (\ref{Fpb}), where the parameter $a$ is estimated from observations.

The metrics or cost functions are  constructed as a mean square of a 3D vector characteristic $\wnu$ of magnetic forces determined at cross-sections $S_\nu$, which in matrix notations is
\begin{eqnarray}
W
&=&
\frac
{1}
{N}
\sum^{\nu_N}_{\nu=\nu_1}
\wnu^{\mathrm T}
\wnu
\, .
	\label{W}
\end{eqnarray}
We have found that two of such characteristics provide the most interesting results.

The first characteristics is derived from Eq. (\ref{f*}) by dividing it on $I$. After normalizing it in the same way as  Eq. (\ref{fnunorm}), one obtains
\begin{eqnarray}
\wnu
&=&
\bfnup
+
C_I
\bfnuIF
\, ,
\qquad
\dbr
{
\frac
{
\Bu
}
{
4\pi
}
}
\, ,
	\label{wnu1}
\end{eqnarray}
which is nothing else than a residual magnetic force per unit current and per unit length of the MFR. Thus, the cost function based on this characteristic is simply a mean square of the effective magnetic field with which the currents of the rope interact.

The second characteristic is derived by dividing Eq. (\ref{wnu1}) on the normalized self-force $C_I \left| \bfnuIF \right|$ to give
\begin{eqnarray}
\wnu
&=&
\frac
{
C_I^{-1}
\bfnup
+
\bfnuIF
}
{
\left|
\bfnuIF
\right|
}
\, .
	\label{wnu2}
\end{eqnarray}
This dimensionless characteristic is a relative residual force with respect to the self-force of the MFR, so that the corresponding cost function is a mean square of this relative force.

Although the absolute minimum $W =0$ is the same for both introduced cost functions, it can ideally be reached only if all $\left|\fnu\right|$, $\nu=\nu_1,\ldots,\nu_N$ vanish during the iterative minimization process of these functions described below.
It turns out, however, that for the used \RBSL\ model of an MFR, the lower bound of $\max\limits_\nu \left|\fnu\right|$ generally does not vanish and depends on the form of the cost function.
Therefore, when starting from the same initial axis path, the minimization of these functions  yields different results.
This raises the important question of how close these results can be made by varying the initial axis path.
We postpone this investigation for the future and consider below in Section \ref{s:examples} only the results for one possible initial axis path.

Note that the parameter $C_I$ ($C_I^{-1}$) enters quadratically into the first (second) $W$, which allows one to find immediately its optimal value for a given axis path. In the first case, we obtain
\begin{eqnarray}
C_I
=
\left. 
-
\sum^{\nu_N}_{\nu=\nu_1}
\bfnup
^{\mathrm T}
\,
\bfnuIF
\right/
\sum^{\nu_N}_{\nu=\nu_1}
\left|
\bfnuIF
\right|^2
	\label{CI1}
\end{eqnarray}
and in the second case
\begin{eqnarray}
C_I
=
\left. 
-
\sum^{\nu_N}_{\nu=\nu_1}
\frac
{
\left|
\bfnup
\right|^2
}
{
\left|
\bfnuIF
\right|^2
}
\right/
\sum^{\nu_N}_{\nu=\nu_1}
\frac
{
\bfnup
^{\mathrm T}
\,
\bfnuIF
}
{
\left|
\bfnuIF
\right|^2
}
\, .
	\label{CI2}
\end{eqnarray}

The optimization of the axis path is a less trivial problem  that generally can be tackled only numerically, because both cost functions have a very complex nonlinear dependence on the coordinates of  control nodes. Therefore, we will solve this numerical problem iteratively in small steps. Let us first perturb  $\wnu$ with small displacements of the nodes (Eq. (\ref{dR})) and linearize it around an unperturbed path to obtain
\begin{eqnarray}
\wnu
\approx
\wnu^{0}
+
\Jnu
\bm{\chi}
\, ,
	\label{wnulin}
\end{eqnarray}
where $\wnu^{0}$ is the unperturbed characteristic and
\begin{eqnarray}
\left(
\Jnu
\right)
_{ij}
\equiv
\pd
{
w^i_\nu
}
{
\chi^{j}
}
	\label{Jnu}
\end{eqnarray}
is a $3\times 2 (N+1)$--dimensional Jacobian matrix determined numerically in terms of Fr\'{e}chet derivatives along the basis vectors $\Mv_j$ and $\Nv_j$, $j=0,\ldots, N$ (Section \ref{s:path}). The substitution of Eq. (\ref{wnulin}) into Eq. (\ref{W}) turns $W$ into a quadratic form in $\bm{\chi}$ with a symmetric and positive definite matrix 
$
\Jnu^{\mathrm T}
\,
\Jnu
$, so that with the minimization of this form we arrive at the classical Gauss--Newton method \citep{Fletcher2000}. This method alone, however, is not sufficient for our purposes, as it may generally result in a $|\bm{\chi}|$ that is too large in value and, therefore, invalidate our linearization approach.

To be self-consistent with this approach, one needs to minimize $W$ subject to the constraint $
\bm{\chi}^{\mathrm T}
\bm{\chi}
=
\mbox{const}
\ll
1
$.
This is reached by extending the cost function as follows:
\begin{eqnarray}
W
&=&
\frac
{1}
{N}
\sum^{\nu_N}_{\nu=\nu_1}
\wnu^{\mathrm T}
\wnu
+
\lambda
\,
\bm{\chi}^{\mathrm T}
\bm{\chi}
\, ,
	\label{Wm}
\end{eqnarray}
where $\lambda$ is a Lagrange multiplier, known in the least-squares method as damping parameter \citep{Levenberg1944, Marquardt1963}. Taking the derivative of this extended cost function with respect to $\bm{\chi}$ and setting the result to zero yields the following linear system of so-called normal equations:
\begin{eqnarray}
\left(
\sum_{\nu=\nu_1}^{\nu_N}
\Jnu^{\mathrm T}
\,
\Jnu
+
\lambda
{\bf I}
\right)
\bm{\chi}
=
-
\sum_{\nu=\nu_1}^{\nu_N}
\Jnu^{\mathrm T}
\wnu^{0}
\, .
	\label{neqs}
\end{eqnarray}
In this form, the derived system is applicable to both cases defined by Eqs. (\ref{wnu1}) and (\ref{wnu2}). However, their Jacobian matrices are different: in the first case 
\begin{eqnarray}
&&
\Jnu
=
\Jnup
+
C_I
\JnuIF
\, ,
	\label{Jnu1}
\end{eqnarray}
where
\begin{eqnarray}
&&
\left(
\Jnup
\right)
_{ij}
\equiv
\pd
{
\fnup^i
}
{
\chi^{j}
}
\, ,
	\label{Jnup}
\\
&&
\left(
\JnuIF
\right)
_{ij}
\equiv
\pd
{
\fnuIF^i
}
{
\chi^{j}
}
\, ,
	\label{JnuIF}
\end{eqnarray}
and in the second case
\begin{eqnarray}
\Jnu
=
\frac
{1}
{
\left|
\bfnuIF
\right|
}
\left(
C_I^{-1}
\Jnup
+
\JnuIF
-
\frac
{
\wnu^{0}
}
{
\left|
\bfnuIF
\right|
}
\,
\bfnuIF^{\mathrm T}
\JnuIF
\right)
\, .
	\label{Jnu2}
\end{eqnarray}

To initialize optimization procedure, we first reconstruct an approximate axis path of the MFR by using observational data and convert it to the canonical form, as described in Section \ref{s:path}. For this canonical path, we compute then the corresponding $\wnu$ and $\Jnu$. By putting $\wnu^{0}=\wnu$ in Eq. (\ref{neqs}) we solve it for $\bm{\chi}$ at several different values of the parameter $\lambda>0$ until the inequality $\max\limits_{i=1,\ldots,2(N+1)} |\chi_i| \ll 1$ is satisfied. We consider that $\bm{\chi}$ satisfying this inequality is an acceptable solution, which we use to calculate by Eqs. (\ref{Rvnu}) and (\ref{dR}) a new axis path $\Rv+\delta\!\Rv$ for the next iteration. We iterate in this manner until $W$ stops decreasing. The canonical path in this sequence of iterates that corresponds to a minimum of $W$ is regarded as a sought-for optimal path.

One should not expect a priori the existence of an axis path that would provide an exact force-free equilibrium. In general, even after all our efforts to reduce the residual force in the constructed PEC, its absolute value will always be above zero. How low the level of this force can be made, with the magnitude of the axial current bounded from below, depends on several factors.

To bring the residual force down by remaining within our approach, one can, in principle, play with the choice of the radius $a$ (1), the initial axis path (2), the form of the cost function (3), and the optimized parameter $C_I$ (4). In Section \ref{s:examples}, we  make the choice of (1) and (2) only once with the help of observational image data.
We explore, however,  what one would obtain by  choosing (3) and (4) in several possible ways.

From the standpoint of the minimization of $W$ only, it would be self-consistent to use at each iterate the expression for $C_I$ derived from the same $W$ as the normal equations. However, if one takes into account the subsequent line-tied zero-beta MHD relaxation of the resulting optimized PEC, this part of the method has to be modified.

It turns out that, for PECs with an ambient potential field of a bipolar type, Eq. (\ref{CI1}) provides somewhat low $C_I$ values, such that the corresponding MFRs are pushed too much toward the solar surface after relaxation and are, therefore,  partially deprived of their initial coherency.
In this respect, the use of Eq. (\ref{CI2}) for $C_I$ leads to better equilibria, where the MFRs hover over the surface, or barely touch it,  as well-defined objects.
This different behavior can be explained in terms of the corresponding expressions for $C_I$.
Note that each of them is a sum of different terms $t_{\nu_{i}}$ over the evaluation nodes, so that $t_{\nu_{i}}/C_I$ is the weight with which the $i$-th node contributes to $C_I$.
The comparison of these weights at different iterations of our optimization process shows that Eq. (\ref{CI1}) generally has a more or less uniform distribution of the weights over the nodes.
In contrast, Eq. (\ref{CI2}) has generally higher weights at the central part of the MFR, where the ambient potential field is stronger.
This implies that a larger value of $|C_I|$ has to be obtained in the latter case, exactly as it is the case in our example study below.
Bearing this in mind, we will  employ only Eq. (\ref{CI2}) from here on, irrespective of which cost function $W$ is used for optimizing the axis paths.

Furthermore, we would like to note that, in general, the radius $a$, the axial current profile, and the corresponding \RBSL-kernels can vary along the MFR.
The right hand side of Eq. (\ref{Fpb}) is then 
not a constant, but a function of $\nu$.
This provides, in principle, additional degrees of freedom for MFR variations, which 
could be used to improve the result of the optimization.
By taking those into account,
however, we would significantly increase the dimensionality of the optimization problem, making it less tractable.
Therefore, we considered here only the simplest \RBSL\ model for MFRs whose diameter and axial current profile do not vary along the axis path.


\section
{
Illustration of how the method works
\label{s:examples}
}
Let us now consider how our improved \RBSL\ method works for relatively simple, yet realistic PECs. As in \cite{Titov2018}, we choose the 2009 February 13 CME event \citep{Patsourakos2009}, whose PEC had an often observed sigmoidal morphology \citep{Miklenic2011}. As in our previous effort, we do not intend here to perfectly reproduce the observed structure. Rather, our aim is to explore the new capabilities of our improved method by applying it to a familiar PEC. By this, we imply the capabilities that arise from the PEC optimization matching the radial component of the photospheric magnetic field.

\begin{figure}[ht!]
\plotone{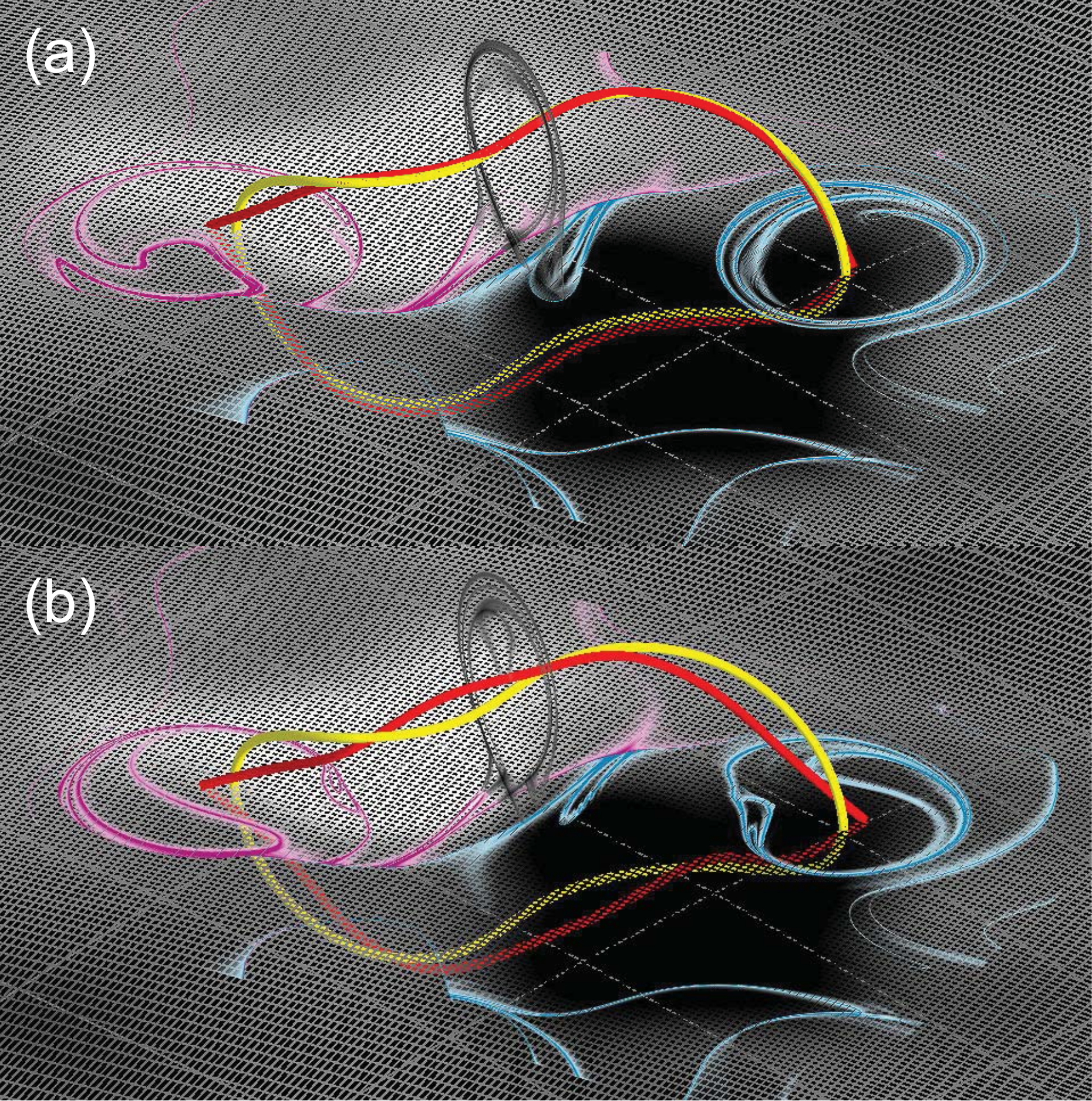}
\caption{
The axis paths  (red lines) and $Q$-maps for optimized Solutions 1 (a) and 2 (b) before MHD relaxation. The yellow line shows, for comparison, the initial axis path, which is the same for both solutions. Only high-$Q$ lines with $\log_{10}Q \ge 4.0$ (sky-blue and crimson for negative and positive polarities, respectively) are shown on top of the photospheric $B_r$-distribution (gray shaded). $Q$-maps are depicted also in the central cross-section of the optimized PECs (inverted grayscale palette with fully transparent colors at $\log_{10}Q<2.0$). The numerical grid is outlined at the boundary. The same color coding is used for similar maps below.
\label{f:ini_paths}}
\end{figure}

\begin{figure*}[ht!]
\plotone{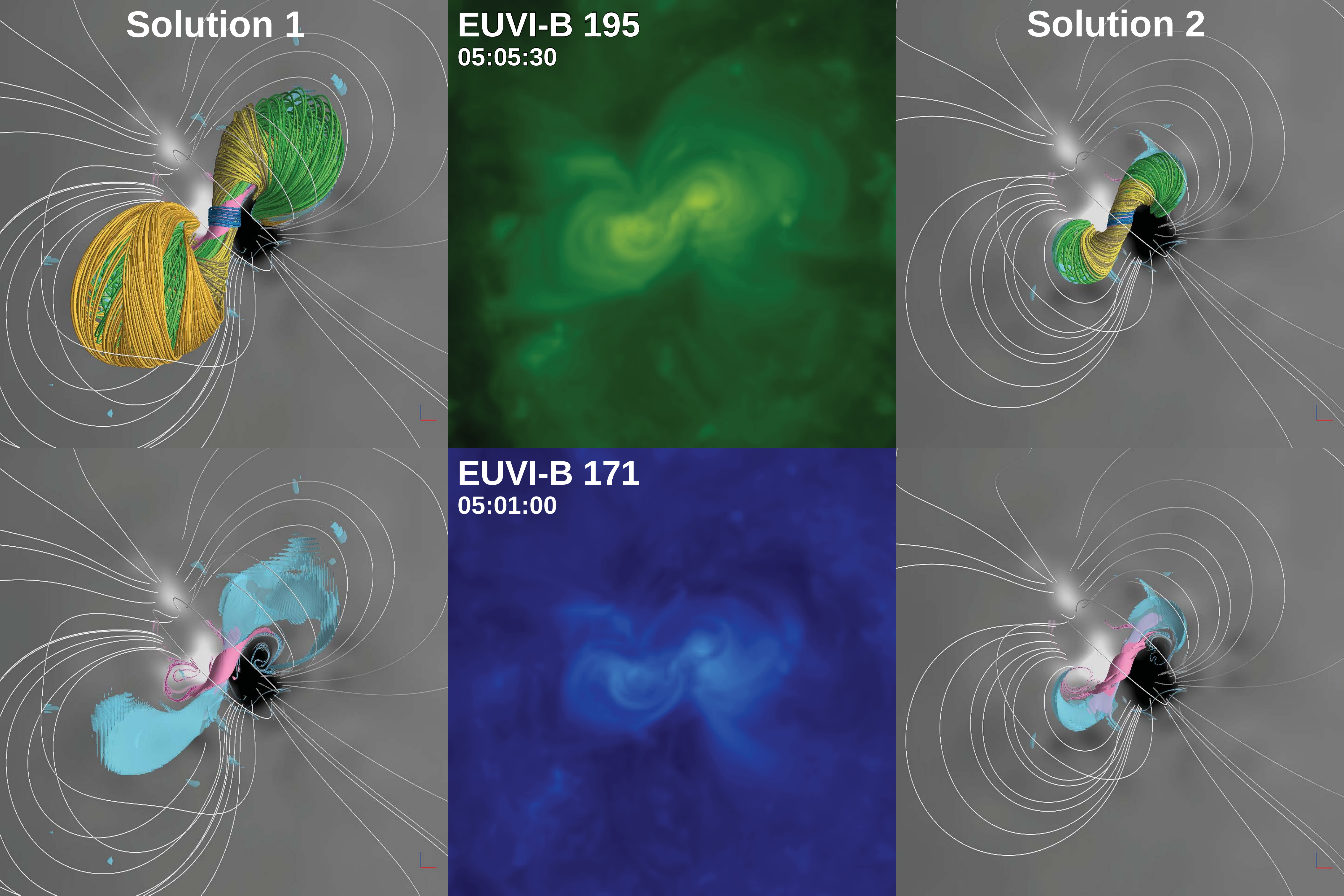}
\caption{
Solutions 1 and 2 vs. STEREO/EUVI EUV images of the PEC of the 2009 February 13 CME: top view on the field-line structure (1st row) and on iso-surfaces of $j/j_{\max}$ = 0.438 (magenta) and $\alpha/\alpha_{\max} = 0.079$ (semi-transparent cyan) (2nd row).
The foot points of field lines in the core field region are chosen for each solution according to the corresponding $Q$-map of the relaxed configuration (see Section \ref{s:Sol1} and \ref{s:Sol2}), so they are somewhat different. Outside of the core region, the same field-line foot points are used for both solutions. 
\label{f:S1_vs_S2_top_view}}
\end{figure*}

As explained in Section \ref{s:opt}, the result of the optimization is not unique and depends on the form of the used cost function $W$. Below we apply two cost functions with $\wnu$ given by Eqs. (\ref{wnu1}) and (\ref{wnu2}), and call the solutions with the corresponding optimized axis paths Solutions 1 and 2, respectively. For both solutions we use the same initial axis path (yellow line in Figure \ref{f:ini_paths}) and $a = 0.01 R_\sun$, where $R_\sun$ is the solar radius. The path was approximated by equidistant control nodes, whose number (nine) was kept unchanged throughout the optimization iterations for both solutions, as the length of the iterated paths did not vary much. In each iteration, we kept the maximal displacement of the control nodes strictly equal to $0.1\, a$. The small displacements allowed us to prevent a deterioration of the path approximation and the associated breakdown of the iterative process (see Sections \ref{s:path} and \ref{s:opt}, respectively).

Both minimums of the cost functions $W$ with $\wnu$ given by Eqs. (\ref{wnu1}) and (\ref{wnu2}) turn out to be relatively shallow.
The value of $W$, calculated in both cases for each iterated axis path, first decreases and then starts to increase with growing number of iterations.
We considered the iteration at which $W$ has a minimum value in this sequence as final, and the corresponding $C_I$ and $\Rv(\nu)$ as optimal.
The minimum of Solution\ 1 (Solution\ 2) is reached at the third (fourth) iteration of the optimization procedure, with the resulting $C_I \simeq -4.12$ ($C_I \simeq -3.72$) and $W$ reduced by $\sim 25\%$ ($\sim 24\%$) relative to its initial value.
We found no evidence for the occurrence of another minimum up to the sixth (eight) iteration.

One reason for the occurrence of shallow minima is the requirement of a constant MFR cross-section in our \RBSL\ formulation (see Section\,\ref{s:AIAF}). We note that, during the subsequent MHD relaxation, the cross sections remain roughly constant in the center of our MFRs  (albeit acquiring an oval shape; see Figures \ref{f:crossec_S1}-\ref{f:case_11c}), while the outer MFR parts expand strongly, especially for Solution 1 (Figures \ref{f:S1_vs_S2_top_view}, \ref{f:FL-structure_S1}, and \ref{f:case_11c}). Thus, even after our optimization, the imbalance between magnetic tension and gradient of magnetic pressure remains large enough to produce such an expansion, which will last until a balance between these components of the magnetic force is reached.

In other words, the preset of constant cross-section of the MFR is generally too restrictive to allow the system to closely approach a force-free state via the optimization of the axial current and the axis path. The MFRs with less expandable legs should be more adaptable to the optimization. Future investigations will show how typical these MFRs are. Fortunately, the subsequent line-tied MHD relaxation can bring even the expandable MFRs close to a force-free state, which significantly extends the applicability of our method.

To analyze the magnetic structures resulting from our MHD relaxations, we calculated  cross-sectional and boundary maps of the squashing degree \citep{Titov2002} or squashing factor $Q$ \citep{Titov2007a} of elementary magnetic flux tubes, which characterizes the divergence of field lines in these tubes. The $Q$-maps helped us to identify the building blocks of those structures whose boundaries are defined by high-$Q$ surfaces. The latter are generally separatrix surfaces, quasi--separatrix layers \citep[QSLs,][]{Priest1995}, or their hybrids. A detailed analysis of the magnetic structure of our solutions is presented in Sections \ref{s:Sol1} and \ref{s:Sol2}, respectively.

The magnetic fluxes of our optimized PECs are partitioned by separatrix surfaces that disappear during the MHD relaxation, thereby yielding their role to newly formed QSLs. The separatrix surfaces are built of coronal field lines that touch the solar surface at the PIL segments called bald patches \citep[BPs,][]{Titov1993}. Such field lines are relatively easy to determine, as we demonstrate in Section \ref{s:Sol1} for Solution 1. The field-line structure of the QSLs can be recovered with the same precision by calculating so-called bracketing field-line pairs \citep[see][]{Titov2017}. While our QSLs possess an intricate internal structure, we restrict ourselves here, for simplicity, to an analysis of the overall magnetic structure, i.e., of the magnetic ``building blocks'' of our relaxed PECs. To do so, we simply identify all flux systems that are separated by QSLs and draw the corresponding field lines with different colors. We also identify all current layers that are present in the system.

The results of this analysis are summarized in Figure \ref{f:S1_vs_S2_top_view}, which compares our two solutions with EUV images of the observed PEC. One can see that both solutions reproduce the observed sigmoidal morphology. However, the MFR in Solution 1 is significantly more inflated, and the MFR in Solution 2 fits the observations better. Note that, despite this significant size difference, the optimized axis paths of Solutions 1 and 2 do not differ much from each other (cf. Figures \ref{f:ini_paths}(a) and \ref{f:ini_paths}(b)), and the difference between the corresponding optimized axial currents is only $\sim 10\%$. Thus, it appears that the result of the MHD relaxation is rather sensitive to the choice of the cost function in the optimization procedure.

One can see from Figure \ref{f:S1_vs_S2_top_view} that the orientation of the modeled MFRs and observed sigmoids noticeably differs from each other. We believe that the main reason for this difference is that the initial MFR footprint locations were not chosen accurately enough in our model. For the purpose of comparison, we chose the same locations as in \citet{Titov2018}, where the observed configuration was modeled with
our earlier version of \RBSL{s}. That version did not preserve the observed magnetogram at the flux-rope footprints, while our upgraded \RBSL\ model used here practically preserves it. Therefore, the previous and present MFRs interact with slightly different ambient potential fields, which likely led to different orientations of the relaxed flux ropes. In other words, the footprint locations that worked well for the previous model are apparently not the best choice for our current model.
The resulting discrepancy could likely be removed by fine-tuning the footprint locations, but this is beyond the scope of the present publication and will be left for a future investigation. 

It is interesting to note that our previous model \citep{Titov2018} used a rather different initial axis path, and it did not invoke an optimization of the MFR parameters or a matching of the radial field at the footprints of the MFR to the observations. Also, in contrast to our present solutions, the axial current $I$ was estimated from the balance of magnetic forces just at one middle point of the MFR axis path, and the contribution of the azimuthal current density $\jF$ to that balance was neglected. Nevertheless, that model was also able to qualitatively reproduce the observed sigmoidal morphology. This implies a certain robustness of the \RBSL\ method in reproducing this morphology regardless of whether the mentioned improvements are used or not.

\subsection{
Solution 1
	\label{s:Sol1}
}

\begin{figure}[t] 
\plotone{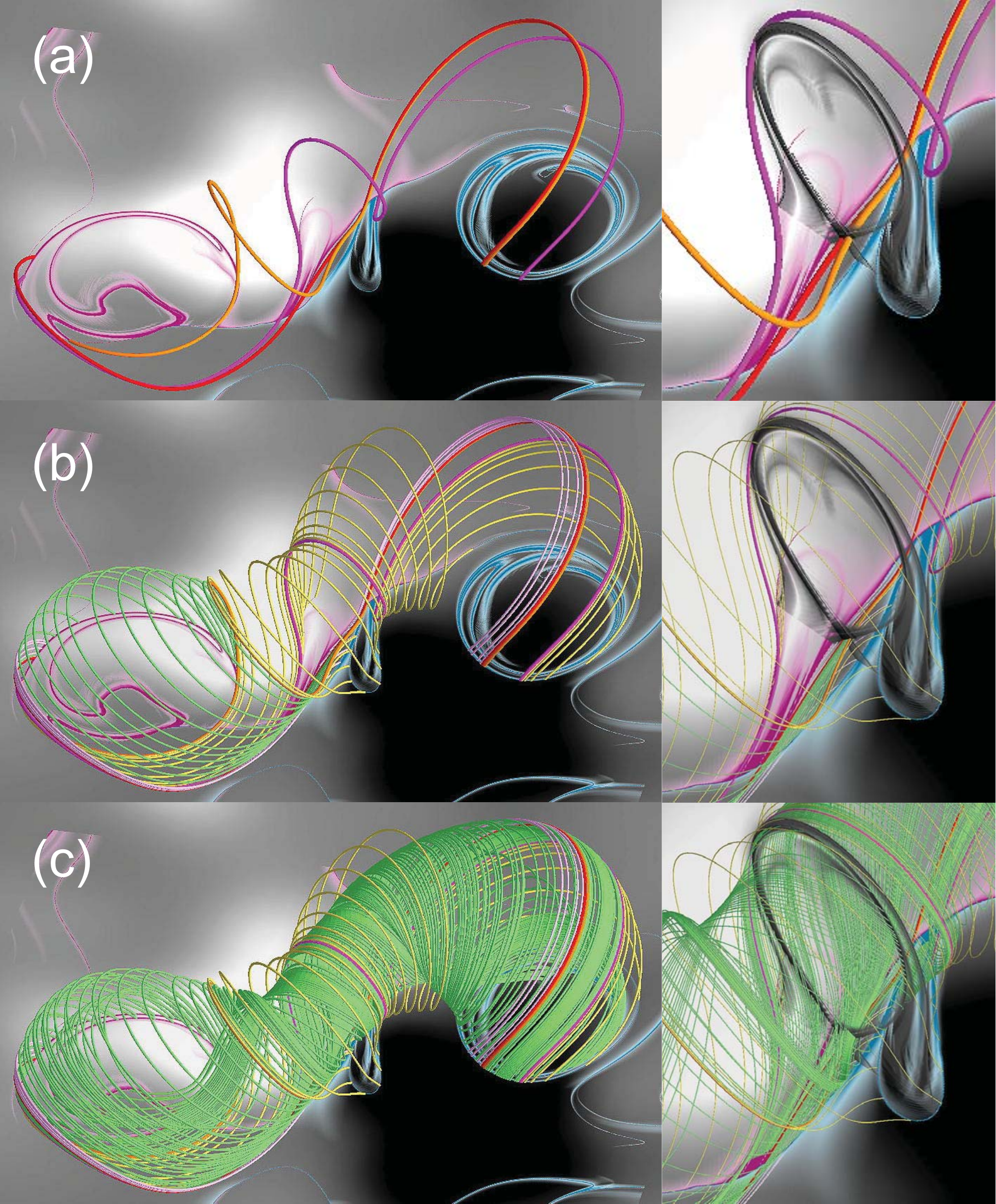}
\caption{
BPSS structure of the optimized PEC for Solution 1, prior to MHD relaxation: (a) BP-BP separators (red, orange, and magenta); (b) BPSS (yellow field lines) that envelopes the MFR, BPSS (light-magenta field lines) that bounds a small arcade below the MFR, and BPSS (green field lines) that fills the gap between two separators (red and orange thick lines) and belongs to the MFR boundary; (c) BPSS (green field lines) that bounds the MFR itself. Panels on the right zoom into the PEC center, to reveal how the field lines are related to the cross-sectional $\log_{10} Q$-map.
\label{f:BPSS-top}}
\end{figure}

\begin{figure*}[ht!]
\plotone{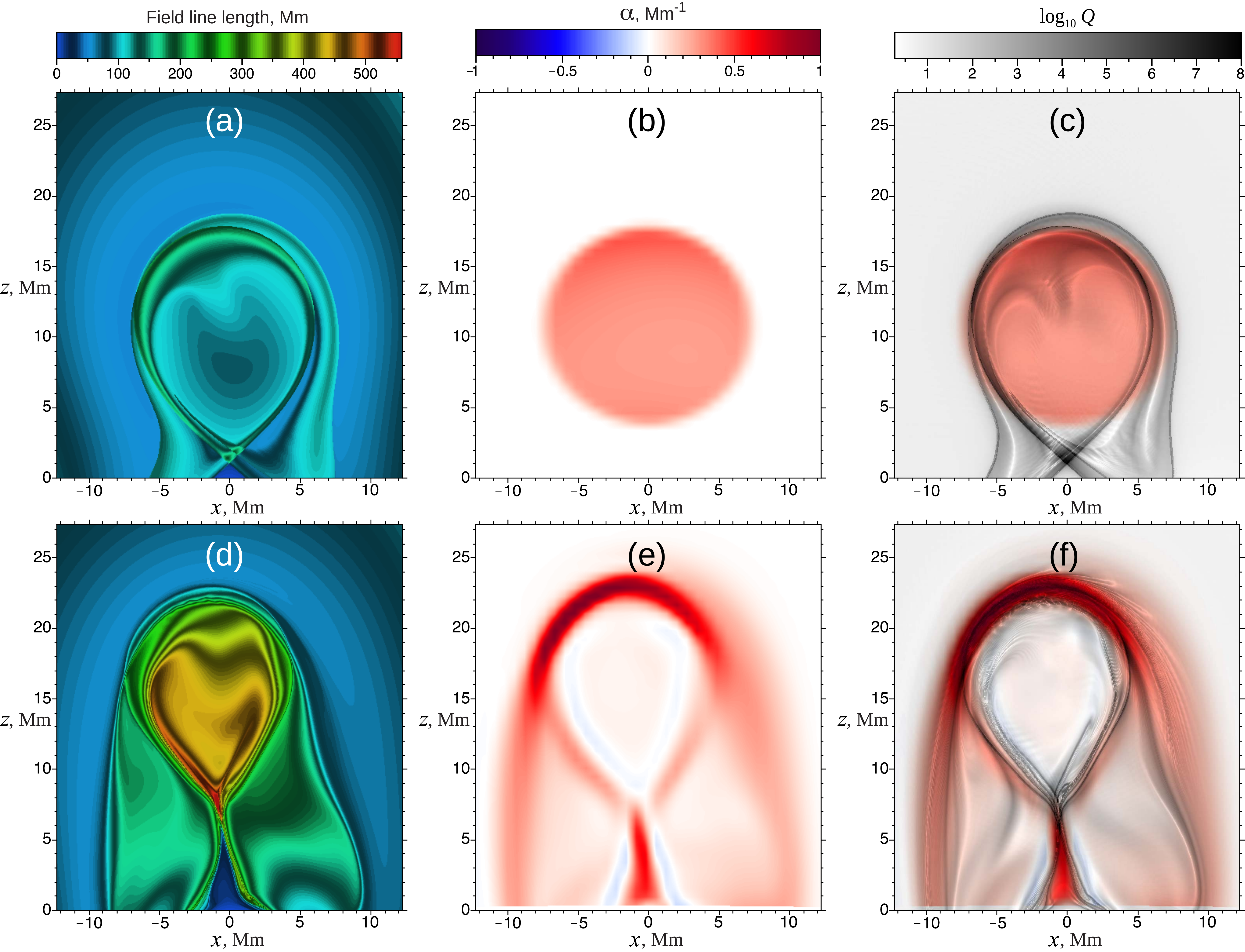}
\caption{
Maps of the field-line length (1st column), force-free parameter $\alpha = (\bm{B} \cdot \bm{\nabla}\times{\bm B}  ) / \bm{B}^2$ (2nd column), and $\log_{10} Q$ (3rd column) in the central PEC cross-section for Solution~1, before (1st row) and after (2nd row) line-tied $\beta=0$ MHD relaxation of the optimized PEC. The greyshaded $\log_{10} Q$-maps are blended with the corresponding blue-red $\alpha$-maps.
\label{f:crossec_S1}}
\end{figure*}

\begin{figure*}[ht!]
\plotone{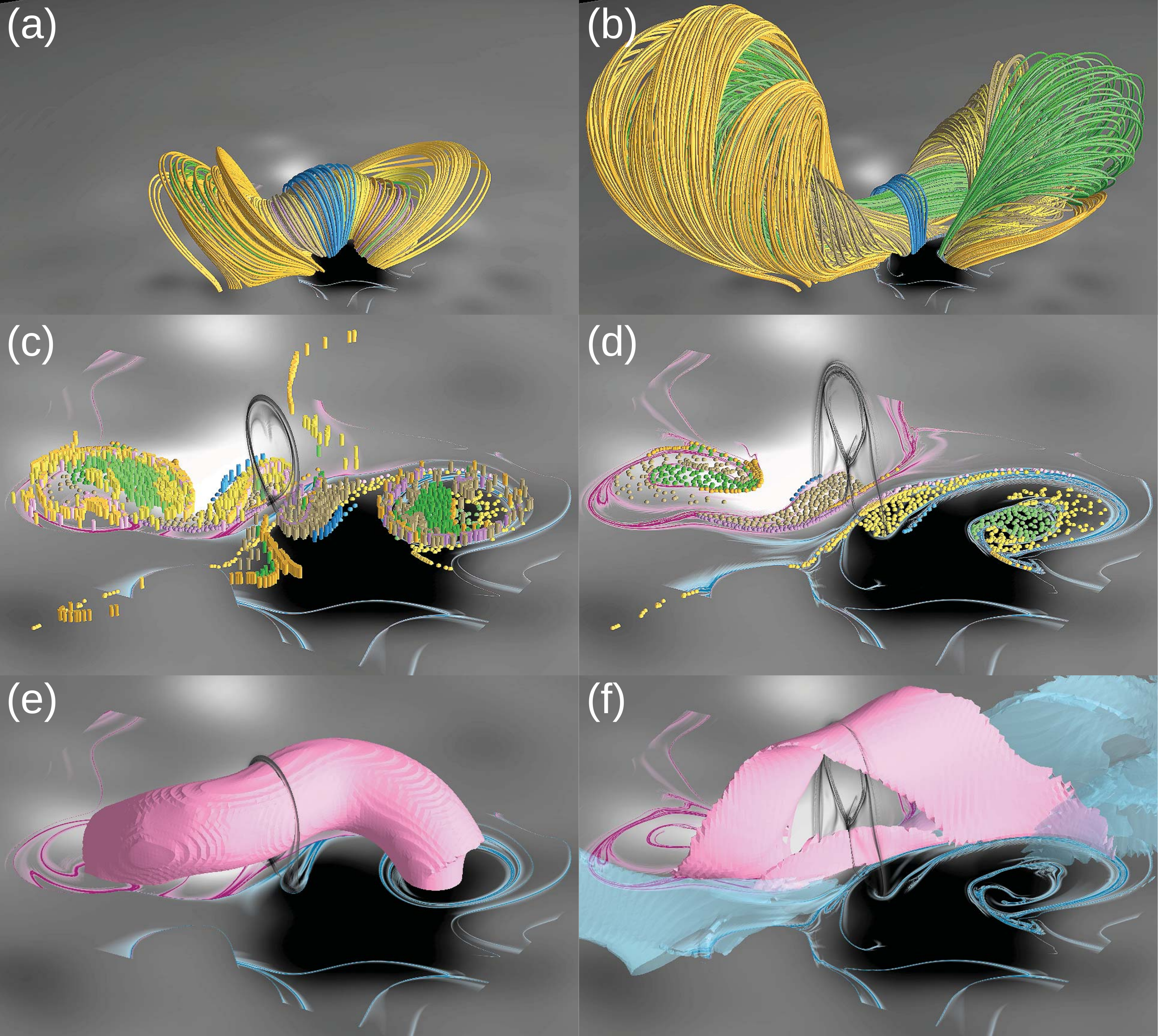}
\caption{
The elements of the PEC before (left column) and after (right column) line-tied $\beta=0$ MHD relaxation for Solution 1. The field lines shown in panels (a) and (b) have the same foot points; the latter are depicted in panels (c) and (d) by small balls, whose conjugate foot points prior to the relaxation are shown in panel (c) by small bars of the same color. Panels (e) and (f) show the iso-surfaces $j/j_\mathrm{max} = 0.438$ of the current density (magenta) before and after the relaxation, respectively. Panel (f) also presents an iso-surface $\alpha/\alpha_\mathrm{max} = 0.079$ of the force-free parameter (semi-transparent cyan), to designate a layer of return current. 
\label{f:FL-structure_S1}}
\end{figure*}

\begin{figure}[ht!]
\plotone{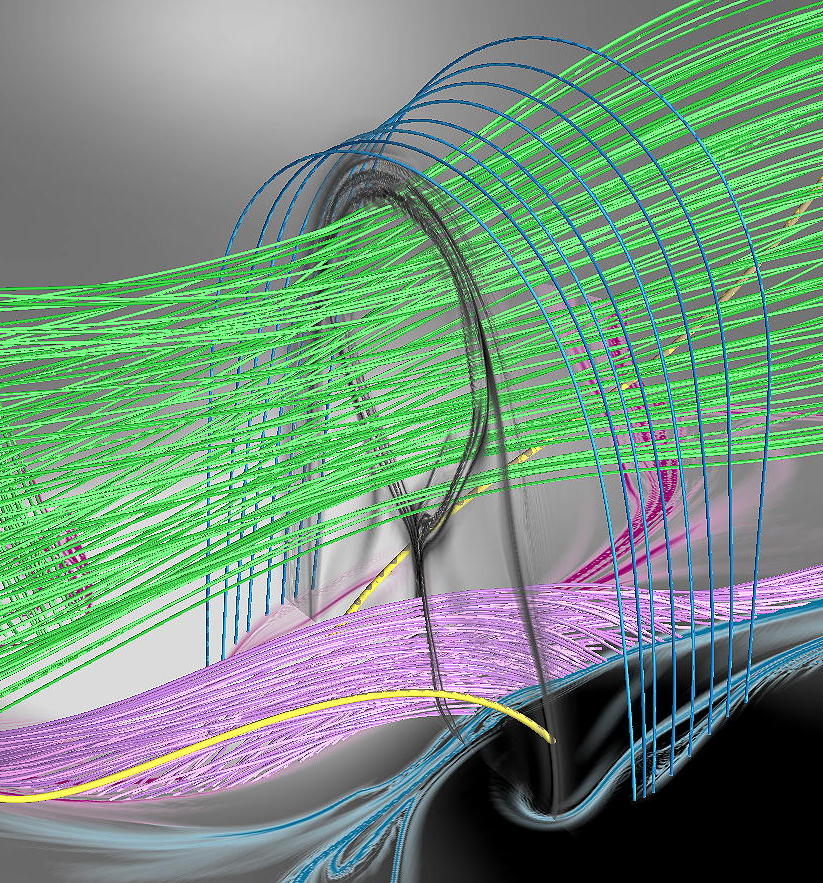}
\caption{
A close-up view onto the relaxed PEC, close to its central cross-section.
Field lines outlining  the enveloping arcade (blue), the MFR (green), the vertical current layer (light-magenta), and two flux tubes adjacent to the current layer (yellow thick lines; called ``arms'' below) are shown. 
\label{f:inlet_CS_MFR_S1}}
\end{figure}

The structural skeleton of the optimized PEC is formed by two bald-patch separatrix surfaces \citep[BPSSs,][]{Titov1993} originating at two segments of the polarity inversion line (PIL) of the photospheric $B_r$-distribution. These BPSSs divide the volume enclosing the MFR into several domains and contain so-called BP-BP separators (red, orange, and magenta thick lines in Figure \ref{f:BPSS-top}), which are the field lines that touch the photosphere twice at BPs and lie at the intersection of two BPSSs. The separator colored in red in Figure \ref{f:BPSS-top} is located below the MFR, and is very similar to the one described for the first time by \citet{Titov1999} in a simple analytical model of a PEC with an arched MFR.

In addition, the structural skeleton of Solution 1 has some topological features that were not covered by that model. The most interesting one is the BPSS that envelopes the MFR boundary and touches it  (see Figures \ref{f:BPSS-top}(b) and \ref{f:crossec_S1}(c)) along the other separator (thick magenta line in Figure \ref{f:BPSS-top}). The appearance of this feature is likely due to the fact that, by construction, the superposition of our MFR and ambient potential fields keeps the photospheric $B_r$-distribution unchanged.

All these BPSSs disappear in the course of the subsequent line-tied MHD relaxation. However, several QSLs, providing a similar partition of the PEC's core magnetic field, are formed during the relaxation. Figure \ref{f:crossec_S1} presents maps of the field-line length $l$, force-free parameter $\alpha$, and $\log_{10} Q$ in the central cross-section of the optimized PEC before (top row) and after (bottom row) the relaxation. By comparing these maps, one can see that the current, which is initially distributed over the whole MFR cross-section, transforms into several force-free current layers, which become aligned with the QSLs. The MFR itself survives this process as a distinct object, which is delineated in the cross-section by high-$Q$ lines. Its teardrop-like shape first shrinks a bit and then substantially rises during the relaxation. Simultaneously, the field lines of the overall MFR increase their length three to five times and acquire an S-like shape (see Figure \ref{f:S1_vs_S2_top_view}).

The cross-sectional $Q$-map in Figure \ref{f:crossec_S1}(c) shows that, prior to relaxation, a small magnetic arcade is present underneath the MFR. In a three-dimensional view, we can see that this arcade is adjoined to the MFR along one of the separators (red) depicted in Figure \ref{f:BPSS-top})(a). As a result of the MFR's rise during the relaxation, the arcade is stretched along the vertical direction, whereas it also develops strong shear and accumulates a large electrical current (see Figures \ref{f:FL-structure_S1}(f) and \ref{f:inlet_CS_MFR_S1}). The development of shear is due to a substantial elongation of the field lines in the core-field region along the horizontal direction (cf. Figure \ref{f:FL-structure_S1}(a) and (b)). Structural features such as this vertical current layer and the adjacent sheared field lines outside it, are generic for many existing models of PECs  \cite[e.g.,][]{Kusano2012, Xia2014a}.

Our $\beta=0$ MHD relaxation was performed under line-tying boundary conditions. Nevertheless, the connectivity of magnetic field lines to the boundary could change due to the presence of (small) resistive and numerical diffusion of the magnetic field. At certain sites of the PECs, however, the connectivity changes were too large for being caused merely by slow diffusion, suggesting that magnetic reconnection took place at those sites. Evidence for that is provided by the above-mentioned disappearance of the initial BPSSs, which is certainly not a small change of the connectivity.

In order to identify such connectivity changes, we plotted field lines in the optimized and relaxed PECs as follows. In both cases, we used for the field lines the same color scheme and launch points. For each of the building blocks of the relaxed PEC, whose cross-sections are outlined in Figure \ref{f:crossec_S1}(f) by high-$Q$ lines, we chose a different color, with a slightly darker (lighter) hue for the field lines launched at the positive (negative) polarity. As launch points we used pairs of conjugate foot points of the relaxed PEC. Any such pair, by definition, gives two identical field lines in the relaxed PEC. However, the corresponding field lines can differ from each other in the optimized PEC, because of the non-vanishing resistivity present in our MHD relaxation. Such non-identical field-line pairs designate the foot points where the magnetic connectivity has changed during the relaxation.

For example, a comparison of Figures \ref{f:FL-structure_S1}(a) and \ref{f:FL-structure_S1} (b) shows that some of the MFR field lines (green) strongly bulged out of the core structure during the relaxation, which hints at a large change of the connectivity at the corresponding foot points. This can be checked by comparing the launch and end points of the paired field lines of the optimized PEC. Depicting the launch and end points by small balls and bars, respectively, one can see that the connectivity significantly changed in the MFR, and also in other building blocks of the core field (cf. Figures \ref{f:FL-structure_S1}(c) and \ref{f:FL-structure_S1}(d)). These changes were apparently caused by magnetic reconnection that was driven in the forming current layers by the residual magnetic stress. By reducing this stress, the reconnection gradually turned into a diffusion at the largely developed current layers, which are shown in Figure \ref{f:FL-structure_S1}(f) for the relaxed PEC.

The set of blue field lines shown in Figure \ref{f:FL-structure_S1}(b) is a magnetic arcade that envelopes the core structure in the center of the PEC. The core contains the sigmodial MFR, which has almost untwisted, but strongly writhed (S-shaped) field lines (green), and two J-shaped magnetic ``loops'' (yellow) that bracket the MFR. In the center of the PEC  (the strong-field region), these loops are nearly horizontal and adjoined to the vertical current layer mentioned above. The loops are much more extended than the enveloping arcade, and, at larger distances from the current layer, wrap around the MFR to add  writhe and sigmoidality to the core structure (see Figures \ref{f:S1_vs_S2_top_view}, \ref{f:FL-structure_S1}(b), and \ref{f:inlet_CS_MFR_S1}).

The vertical current layer underneath the MFR is composed of three sub-layers (see Figure \ref{f:crossec_S1}(f)). The central sub-layer is a narrow sheared arcade that consists of relatively short field lines (light-magenta in Figure \ref{f:inlet_CS_MFR_S1}), which are aligned with the PIL. The adjacent two sub-layers contain much longer field lines, which arch above the MFR and the current layer. One set of foot points of these field lines resides next to the sheared arcade, while the other is located far away from the PIL, at the outskirts of the conjugate polarity. These field lines, colored in orange in Figure \ref{f:FL-structure_S1}(b), have shapes that are similar to the neighboring yellow ones, but they interlock slightly differently with the MFR field lines (green).

\subsection{
Solution 2
	\label{s:Sol2}
}

We performed an MHD relaxation also for Solution 2, which has a weaker optimized axial current than Solution 1 ($C_I$ is about 10\,\% smaller; see above). The relaxation of this case was accompanied by magnetic reconnection as well, and it resulted in a similar equilibrium structure.

\begin{figure*}[ht!]
\plotone{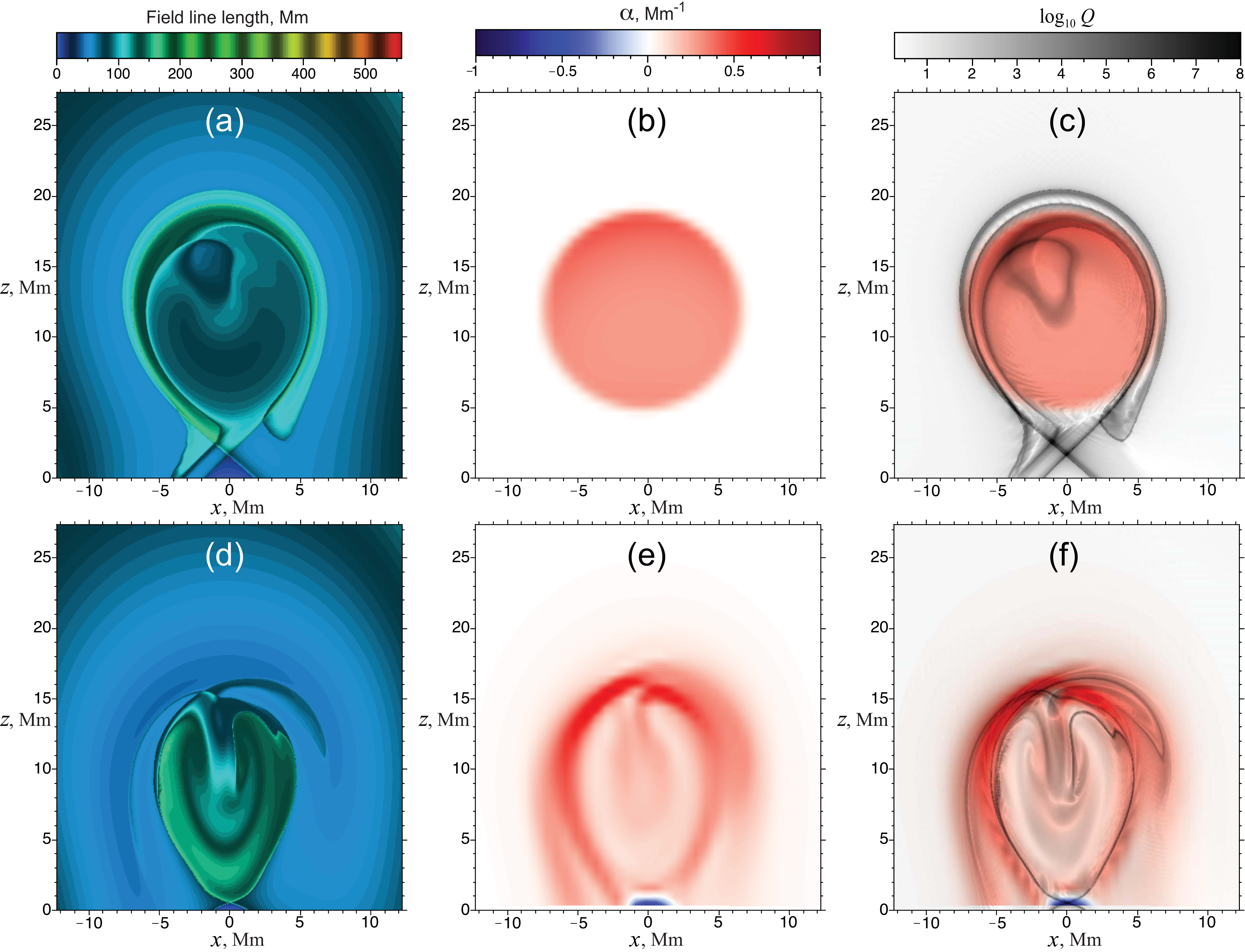}
\caption{
Maps of the field-line length (1st column), $\alpha$ (2nd column) and $\log_{10} Q$ (3rd column) in the central cross-section for Solution~2 before (1st row) and after (2nd row) line-tied $\beta=0$ MHD relaxation of the optimized PEC. The grayshaded $\log_{10} Q$-maps are blended with the corresponding blue-red $\alpha$-maps. The color-bar scales are the same as in Figure \ref{f:crossec_S1}.
}
\label{f:crossec_S2}
\end{figure*}

Figure \ref{f:crossec_S2} shows maps of the field-line length, force-free parameter $\alpha$, and of $\log_{10} Q$ in the central cross-section of the optimized PEC,  before and after the relaxation. Just as for Solution 1, the initially diffuse distribution of the current density transforms into several relatively sharp layers, which are largely aligned with the QSLs that form during the relaxation. Comparing  Figures \ref{f:crossec_S1}(d) and \ref{f:crossec_S2}(d), one can see that the field lines of the relaxed MFR are much shorter than in Solution 1, yielding a much more compact core field (Figure \ref{f:case_11c}(b); see also Figure \ref{f:S1_vs_S2_top_view}). Note that, due to the weaker current, the MFR center is pushed downwards (rather than upwards) in this case. This results in the formation of a horizontal (rather than a vertical) current layer, in which the current flows in the opposite direction (see Figures \ref{f:crossec_S2}(f) and \ref{f:case_11c}(a)-(c)). The formation of this current layer could be prevented by a suitable increase of $|C_I|$ (as in Solution 1). We note that additional adjustments of the modeled PEC could also be obtained by modifying, for example, the shape of the initial axis path or the location of its foot points.

\subsection{
Concluding Remarks
	\label{s:CR}
}

It should be emphasized that the magnetic structure of our specific Solutions 1 and 2 is probably representative for many sigmoidal PECs on the Sun. For example, the field lines shown in Figure \ref{f:case_11c}(d) can readily be associated with the ``envelope'', ``elbows'', and ``arms'' typically seen in observations of PECs in bipolar active regions \citep{Moore2001}. In our solutions, the electrical current is concentrated in relatively thin current layers, which are aligned with QSLs and reach the photospheric boundary along segments of high-$Q$ lines that are located close to the PIL.

It has yet to be verified how generic such a current structure is for sigmoidal PECs, but intuitively it seems that the layers constitute an inevitable interface between flux tubes of different types of geometry and connectivity, as described above. If this is indeed the case, then our solutions demonstrate the unique capability of the improved \RBSL\ method to model these kind of equilibria. Given the complex pattern of QSL footprints and current densities at the boundary, and the limitations of other methods discussed in Section\,\ref{intro}, we do not believe that such PECs can be reproduced in a simple manner by either of those methods.

In contrast, these complex structures are easily obtained by our improved \RBSL\ method from relatively simple input data. The input includes only the photospheric distribution of the normal component of the magnetic field, and an approximate axis path and diameter of the MFR to be modeled. The optimization adjusts the path and provides an approximate value of the axial current required for keeping the optimized MFR in equilibrium. From a topological point of view, this procedure corrects the (quasi-)separatrix surfaces that partition the magnetic flux of the PEC. It is important here that this correction stems from the minimization of unbalanced magnetic forces in the PEC{, so that the subsequent MHD relaxation of the resulting imbalance can only minimally affect the established partition of the magnetic flux.
During this step, the field-line connectivity is largely preserved due to the line-tying conditions imposed at the photospheric boundary. It changes only at the current layers that are self-consistently formed near (quasi-)separatrix surfaces. These connectivity changes are produced by magnetic reconnection, which is driven by not yet balanced magnetic forces. As soon as a force balance is reached in the PEC, the reconnection ceases and only slow magnetic diffusion commences in the current layers.

In other words, the complex current structure in our solutions is a result of the self-organization of the configuration during its MHD relaxation. The role of the boundary during this process is merely to preserve the bulk of the magnetic connectivity that was approximated during the previous step. In this respect, our improved \RBSL\ method is very similar to the MFR insertion method by \citet{vanBall2004}. However, our method has the essential advantage that it makes the magnetic connectivity of the pre-relaxed PEC more adequate to the MFR shape suggested by the observations. This significantly reduces, or even fully eliminates, the number of subsequent trial-and-error relaxation attempts. This advantage of our method is achieved by minimizing the imbalance between magnetic tension and magnetic-pressure gradients in the MFR prior to the MHD relaxation.

\begin{figure*}[ht!]
\plotone{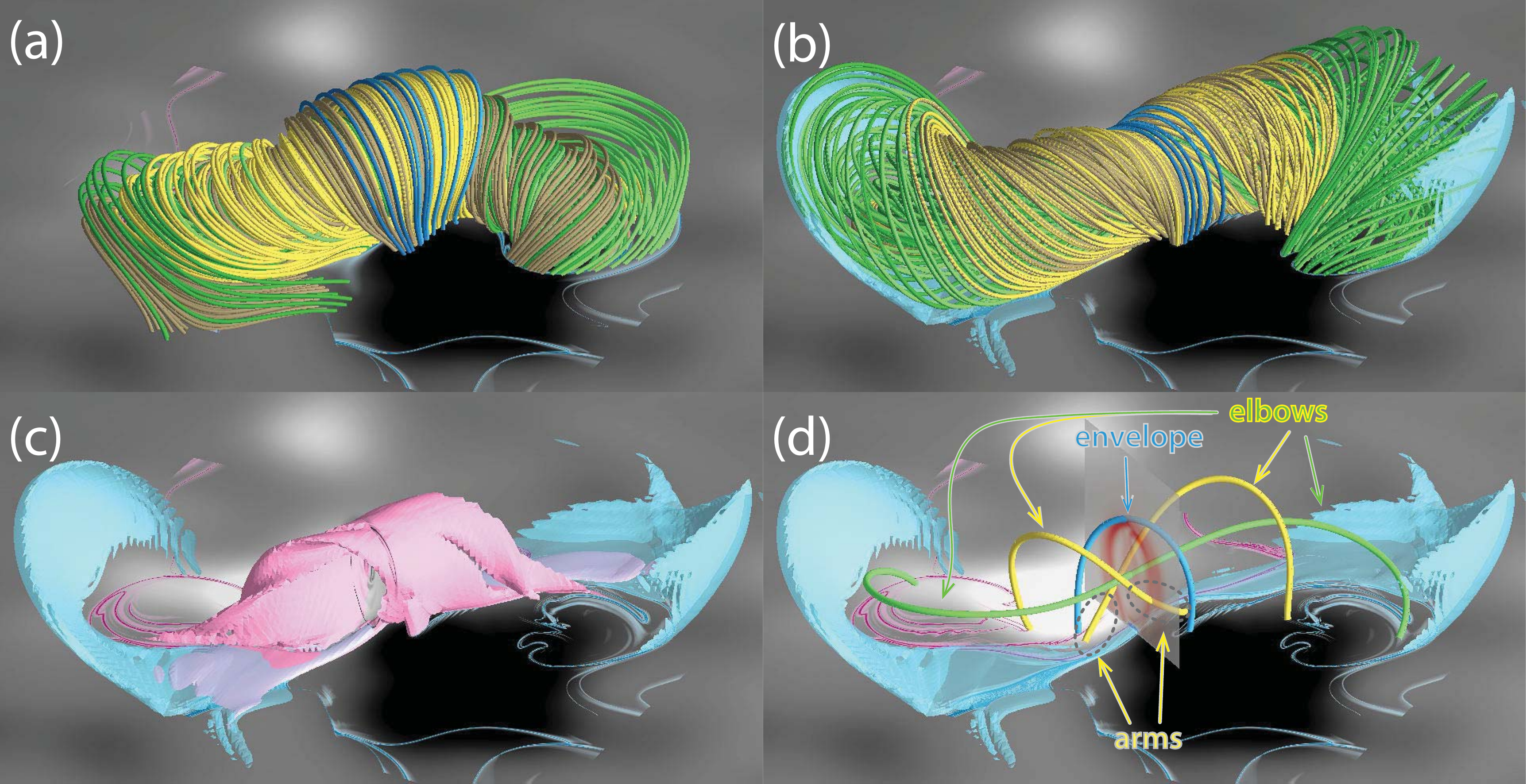}
\caption{
The magnetic structure of 
Solution 2. (a) The \RBSL\ MFR whose parameters minimize $W$ with $\wnu$ and $C_I$ defined by Eqs. (\ref{wnu2}) and (\ref{CI2}), respectively;
(b) the same structure after line-tied zero-beta MHD relaxation;
(c) iso-surface $j/j_{\mathrm{max}} =  0.438$ (magenta) of the current density and the iso-surface $\alpha/\alpha_{\mathrm{max}} = 0.079$ (semi-transparent cyan) of the force-free parameter to designate the corresponding layers of direct and return currents;
(d) three different types of field lines that form the resulting structure.
Semi-transparent iso-surfaces of $|\bm{j}|$ and $\alpha\ (<0)$ are colored in magenta and cyan, respectively.
Panel (d) presents also a $|\bm{j}|$--distribution in the middle cross-section of the PEC.
The photospheric $B_r$--distribution is shown by grey shading from white ($B_r>0$) to black ($B_r<0$). 
	\label{f:case_11c}}
\end{figure*}


\section{Summary}
\label{sum}

We have improved our \RBSL\ method \citep{Titov2018} for modeling PECs, by extending it in two ways. First, we have modified the method so that it allows one to construct, in a straightforward manner, an MFR field with a vanishing or negligibly small normal component at the photospheric boundary. This modification is particularly valuable at the locations of the MFR footprints, where the original method required a more complicated approach for preserving the photospheric normal component of the background field. The perturbation of that component by the insertion of the MFR is now only due to the curvature of the solar surface and, therefore, negligible if the distance between the footprints of the MFR is much less than the solar radius. 

Second, we have developed an optimization method for minimizing unbalanced residual magnetic forces prior to the MHD relaxation of a modeled PEC. This minimization is obtained by optimizing the shape and axial current of the corresponding MFR with the Gauss-Newton method of least squares. To give an idea of how the method performs in practice, we note that our present implementation (written in Maple and Fortran) allows one to conduct an optimization for cases such as those shown in Section\ \ref{s:examples} on a laptop computer within less than one hour. The performance can be improved by fully implementing the optimization method in Fortran.

Our improved \RBSL\ method allows one not only to minimize residual magnetic forces, but also to properly adjust the magnetic connectivity in PEC configurations. In order to evaluate the power of these new capabilities, we combined the method with line-tied $\beta=0$ MHD simulations to construct two numerically relaxed, approximately force-free PEC solutions for the 2009 February 13 CME event, which we had used previously for testing the original model \citep{Titov2018}. The main outcomes of this evaluation can be summarized as follows.

The MFR, as the current-carrying entity in our two optimized PECs has, by construction, a curved cylindrical body whose diameter  is essentially the same over its length. However, during the relaxation, it acquires the shape of a sigmoid with inflated elbows. This transformation is accompanied by a change of the current density distribution, and by the conversion of twist into writhe and shear. The resulting relaxed PECs have a complex core magnetic structure, with the MFR nested within a sheared magnetic arcade. Both the MFR and the arcade are bounded in the central region of the PECs by curved current layers, where the newly developed shear is concentrated.

Depending on the strength of the axial current in the  pre-relaxed MFR, the core of the final PEC can also contain a vertical current layer, which is  then embedded in the sheared arcade, underneath the MFR. This vertical current layer itself is just another, lower-lying sheared arcade whose field lines are much shorter than the adjacent core field lines.

It is interesting to note that all these current layers are well aligned with QSLs that form during the MHD relaxation. The partition of the core field by the QSLs reveals building blocks that match the morphological features typically observed in bipolar pre-eruptive configurations \citep[e.g.,][]{Moore2001} very well.

We believe that this agreement is not just a coincidence, but rather a result of the increased accuracy of our improved \RBSL\ method in constructing approximate magnetic equilibria. This suggests that the method will not only be beneficial as a tool for modeling solar eruptions, but also for scientific studies that require a detailed understanding of the magnetic structure of PECs.


\acknowledgments

We would like to thank the anonymous referee for his/her quick response and thorough report.

This research was supported by NASA programs HTMS (award\,no.\,80NSSC20K1274), HSR (80NSSC19K0858 and 80NSSC20K1317), SBIR (80NSSC19C0193); NASA/NSF program DRIVE (80NSSC20K0604); NSF grants AGS-1560411, AGS-1135432, AGS-1923377, ICER-1854790; and AFOSR contract FA9550-15-C-0001. Computational resources were provided by NSF's XSEDE and NASA's NAS facilities.


\newpage
\appendix
%
\section{Integral \RBSL\ formulations}
\label{s:int_RBSL-forms}

For efficient and accurate calculation of the line density of magnetic force defined by Eq. (\ref{f*}), it is useful to derive the \RBSL{s} that explicitly define the magnetic field and current density associated with our MFR. By using Eqs. (\ref{BI}) and (\ref{AI}) one can straightforwardly obtain the azimuthal component of this field
\begin{eqnarray}
\BI
&=&
\int_{\mathcal{C}\, \cup\,\Cst}
K_{B_I}(r)
\,
\Rvp
\times
\rv
\;
\frac{\dl}{a}
\, ,
\qquad
\dbr
{
\frac
{\mu I}
{4\pi a}
}
\, ,
	\label{BI_int}
\end{eqnarray}
where
\begin{eqnarray}
K_{B_I}(r)
&=&
-
\frac
{
K_I^\prime(r)
}
{
r
}
=
\begin{cases}
\frac
{16}
{3\pi} 
\approx
1.698
\, ,
& \text{$r = 0^{+}$,}
\\
\frac
{2}
{\pi} 
\left[
\frac
{1}
{r^2}
\left(  
\frac
{\arcsin r}
{r}
-
\sqrt{1-r^2}
\right)
+
2
\,
\sqrt{1-r^2} 
\right]
\, ,
& \text{$r \in (0,1]$,}
\\
r^{-3}
\, ,
& \text{$r>1$.}
\end{cases}
	\label{KBI}
\end{eqnarray}

Taking curl of Eq. (\ref{BI_int}), we similarly obtain the axial current density
\begin{eqnarray}
\jI
&=&
\int_{\mathcal{C}\, \cup\,\Cst}
\left[
K_{j_{I_1}}(r)
\,
\Rvp
+
K_{j_{I_2}}(r)
\,
\left(
\Rvp 
\bcd
\rv
\right)
\rv
\right]
\frac{\dl}
   {a}
\, ,
\qquad
\dbr
{
\frac
{I}
{4\pi a^2}
}
\, ,
	\label{jI}
\end{eqnarray}
where the corresponding two kernels are given by
\begin{eqnarray}
K_{j_{I_1}}(r)
&=&
\frac{\left(
      r^2\, K_{B_I}(r)
      \right)^\prime
     }
     {r}
=
\begin{cases}
\frac
{32}
{3\,\pi}
\approx
3.395
\, ,
& \text{$r = 0^{+}$,}
\\
\frac
{2}
{\pi}
\left[
\frac
{1}
{r^2}
\left(
\sqrt{1-r^2}
-
\frac
{\arcsin r}
{r}
\right)
+
6
\sqrt{1-r^2}
\right]
  \, ,
& \text{$r \in (0,1]$,}

\\
-
r^{-3}
\, ,
& \text{$r>1$,}
\end{cases}
	\label{KjI1}
\\
K_{j_{I_2}}(r)
&=&
-
\frac{K^\prime_{B_I}(r)}{r}
=
\begin{cases}
\frac
{16}
{5\,\pi}
\approx
1.019
\, ,
& \text{$r = 0^{+}$,}
\\
\frac{2}{\pi\,r^4}
\left[
3
\frac
{\arcsin r}
{r}
-
\left(
2r^2+3
\right)
\sqrt{1-r^2}
\right]
\, ,
& \text{$r \in (0,1]$,}

\\
3\,r^{-5}
\, ,
& \text{$r>1$.}
\end{cases}
	\label{KjI2}
\end{eqnarray}

The other components of the field and current density associated with the axial flux $F$ are derived in a similar way such that Eqs. (\ref{BF}) and (\ref{AF}) yield the axial magnetic field
\begin{eqnarray}
\BF
&=&
\int_{\mathcal{C}}
\left[
K_{B_{F_1}}(r)
\,
\Rvp
+
K_{B_{F_2}}(r)
\,
\left(
\Rvp
\bcd
\rv
\right)
\rv
\right]
\frac{\dl}{a}
-
\int_{\mathcal{\Cst}}
\left[
\ldots
\right]
\frac{\dl}{a}
\, , 
\qquad
\dbr
{
\frac
{F}
{4\pi a^2}
}
\, ,
	\label{BF_int}
\end{eqnarray}
where $[\ldots]$ represents the same sub-integral expression as in the first integral taken over the path $\mathcal{C}$ and the corresponding kernels are
\begin{eqnarray}
K_{B_{F_1}}(r)
&=&
\frac
{\left(r^2 K_F(r)
 \right)^\prime}
{r}
=
\begin{cases}
\frac
{5}
{\sqrt{6}}
+
\frac
{10}
{\pi}
\left[
\frac{2}{3}
-
\frac
{1}
{\sqrt{6}}
\arcsin
\left(
\frac
{1}
{5}
\right)
\right]
\approx
3.902
\, ,
& \text{$r = 0^{+}$,}
\\
\frac
{2}
{
\pi
r^2
} 
\left(
\sqrt
{
1
-
r^2
}
-
\frac
{
\arcsin r
}
{r}
\right)
+
\frac
{8}
{\pi}
\sqrt{1-r^2}
\\
+
\frac
{1}
{\sqrt{6}}
\left(
5
-
4\,
r^2
\right)
\left[
1
-
\frac
{2}
{\pi}
\arcsin
\left(
\frac
{
1
+
2\,
r^2
}
{
5
-
2\,
r^2
}
\right)
\right]
\, ,
& \text{$r \in (0,1]$,}

\\
-
r^{-3}
\, ,
& \text{$r>1$,}
\end{cases}
	\label{KBF1}
\\
K_{B_{F_2}}(r)
&=&
-
\frac{K_F^\prime(r)}{r}
=
\begin{cases}
\frac
{\sqrt{6}}
{3}
+
\frac
{2}
{
15\,
\pi
}
\left[
24
-
5
\sqrt{6}
\arcsin
\left(
\frac
{1}
{5}
\right)
\right]
\approx
1.730
\, ,
& \text{$r = 0^{+}$,}
\\
\frac
{2}
{\pi r^4} 
\left[
3
\,
\frac
{\arcsin r}
{r}
-
\left(
3
+
2\,
r^2
\right)
\sqrt{1-r^2}
\right]
\\
+
\frac
{2}
{\sqrt{6}}
\left[
1
-
\frac
{2}
{\pi}
\arcsin
\left(
\frac
{
1
+
2\,
r^2
}
{
5
-
2\,
r^2}
\right)
\right]
\, ,
& \text{$r \in (0,1]$,}

\\
3\,r^{-5}
\, ,
& \text{$r>1$.}
\end{cases}
	\label{KBF2}
\end{eqnarray}

The curl of Eq. (\ref{BF_int}) gives the respective azimuthal current density
\begin{eqnarray}
\jF
&=&
\int_{\mathcal{C}}
K_{j_F}(r)
\,
\Rvp
\times
\rv
\;
\frac{\dl}{a}
-
\int_{\Cst}
[\ldots]\;
\frac{\dl}{a}
\, , 
\qquad
\dbr
{
\frac
{F}
{4\pi\mu a^3}
}
\, ,
	\label{jF_int}
\end{eqnarray}
where $[\ldots]$ represents the same sub-integral expression as in the first integral taken over the path $\mathcal{C}$, and the corresponding kernel is
\begin{eqnarray}
K_{j_F}(r)
=
-
\frac{\left(
      r^4 K_F^\prime(r)
      \right)^\prime
     }
     {r^4}
=
\begin{cases}
\frac
{16}
{\pi}
+
\frac
{10}
{\sqrt{6}}
\left[
1
-
\frac
{2}
{\pi}
\arcsin
\left(
\frac
{1}
{5}
\right)
\right]
\approx
8.652
\, ,
& \text{$r = 0^{+}$,}
\\
\frac
{
5
\sqrt{6}
}
{3}
+
\frac
{40}
{\pi}
\frac
{
2
-
r^2
}
{
(
5
-
2\,
r^2
)
\sqrt{1-r^2}
}
-
\frac
{20}
{
\sqrt{6}\,
\pi
}
\arcsin
\left(
\frac
{
1
+
2\,
r^2
}
{
5
-
2\,
r^2
}
\right)
\, ,
& \text{$r \in (0,1)$,}

\\
0
\, ,
& \text{$r>1$.}
\end{cases}
 	\label{KjF}
\end{eqnarray}
This kernel diverges as $ (1-r)^{-1/2} $ at $r = 1$, so it is integrable. Therefore, the integral given by Eq. (\ref{jF_int}) is  defined and, in principle, can be calculated by  using  a suitable change of variables in the neighborhood of the singularity. However, before making such a calculation, the singular points have to be first localized and then bracketed on the path, which generally might be needed to repeat several times depending on the shape of the path and the point $\rv$ at which $\jF$ is determined. Therefore, the implementation of this calculation of $\jF$ is not a simple task, since it requires the use of a complex logic and exception handling.

For this reason, we implemented a technically simpler method that determines $\jF$ by taking numerically curl of $\BF$ with finite differencing. As the kernels defined by Eqs. (\ref{KBF1}) and (\ref{KBF2}) have no singularities, the required values of $\BF$ for such differencing straightforwardly follow from Eq. (\ref{BF_int}).

\section{Differential \RBSL\ formulations
\label{s:diff_RBSL-forms}
}

We have found that the computation of the vector potentials ${\bm A}_I$ and ${\bm A}_F$ as well as the MFR fields and current densities described in Appendix \ref{s:int_RBSL-forms} can be made much more efficient if we represent  the corresponding \RBSL\ integrals in terms of solutions of certain ODEs. Indeed, regarding $\xv$ as a fixed parameter and changing the variable of integration in Eq. (\ref{AI}) from arc length $l$ to parameter $\nu$ (see its definition in Section \ref{s:path}), we obtain that ${\bm A}_I(\nu,\xv)$ satisfies the following ODE:
\begin{eqnarray}
\frac
{\mathrm{d}\ }
{\mathrm{d}\nu}
{\bm A}_I(\nu,\xv)
  &=&
\frac
{1}
{a}
K_I(r)
\;
\Rvd
\, , 
\qquad
\dbr
{
\frac
{\mu I}
{4\pi}
}
\, ,
	\label{AIdiff}
\end{eqnarray}
in which $r=|\rv|$ and
$$
\rv
\equiv
\rv(\nu)
=
\frac
{\xv -\Rv(\nu)}
{a}
\, .
$$
By supplementing also this equation with Eqs. (\ref{Rvnu}), (\ref{dRvnu}), thereby defining the axis path, and the initial condition
\begin{eqnarray*}
{\bm A}_I(0,\xv) = {\bm 0}
\, ,
	\label{AI0}
\end{eqnarray*}
we set the initial value problem for this ODE such that its solution evaluated at $\nu=N$, which is ${\bm A}_I(N,\xv)$, will provide the integral over the coronal path $\mathcal{C}$ in Eq. (\ref{AI}). 
To obtain the remaining integral over the subphotospheric path $\Cst$, one needs first to change  ${\bm A}_I$ for ${\bm A}^{*}_I$, $\Rv(\nu)$ for $\Rv^{*}(\nu)$, and $\Rvd(\nu)$ for $\Rvdst(\nu)$ in Eq. (\ref{AIdiff}), and then $-{\bm A}^{*}_I(N,\xv)$ will determine the value of this integral. The negative sign here takes into account the proper direction of the integration over the path $\Cst$. 
The differential formulations for the \RBSL{s} given by Eqs. (\ref{BI_int}), (\ref{jI}), (\ref{AF}), (\ref{BF_int}), and (\ref{jF_int}) are obtained in a similar way to give
\begin{eqnarray}
\frac
{\mathrm{d}\ }
{\mathrm{d}\nu}
\BI(\nu,\xv)
&=&
\frac{1}{a}
K_{B_I}(r)
\,
\Rvd
\times
\rv
\, ,
\qquad
\dbr
{
\frac
{\mu I}
{4\pi a}
}
\, ,
	\label{BIdiff}
\\
\frac
{\mathrm{d}\ }
{\mathrm{d}\nu}
\jI(\nu,\xv)
&=&
\frac{1}{a}
\left[
K_{j_{I_1}}(r)
\,
\Rvd
+
K_{j_{I_2}}(r)
\,
\left(
\Rvd 
\bcd
\rv
\right)
\rv
\right]
\, ,
\qquad
\dbr
{
\frac
{I}
{4\pi a^2}
}
\, ,
	\label{jIdiff}
%
\\
\frac
{\mathrm{d}\ }
{\mathrm{d}\nu}
{\bm A}_F(\nu,\xv)
&=&
\frac
{1}
{a}
K_F(r)
\,
\Rvd
\times
\rv
\, , 
\qquad
\dbr
{
\frac
{F}
{4\pi a}
}
\, ,
	\label{AFdiff}
%
\\
\frac
{\mathrm{d}\ }
{\mathrm{d}\nu}
\BF(\nu,\xv)
&=&
\frac{1}{a}
\left[
K_{B_{F_1}}(r)
\,
\Rvd
+
K_{B_{F_2}}(r)
\,
\left(
\Rvd
\bcd
\rv
\right)
\rv
\right]
\, , 
\qquad
\dbr
{
\frac
{F}
{4\pi a^2}
}
\, ,
	\label{BFdiff}
%
\\
\frac
{\mathrm{d}\ }
{\mathrm{d}\nu}
\jF(\nu,\xv)
&=&
\frac{1}{a}
K_{j_F}(r)
\,
\Rvd
\times
\rv
\, , 
\qquad
\dbr
{
\frac
{F}
{4\pi\mu a^3}
}
\, .
	\label{jBdiff}
\end{eqnarray}

To calculate the correct total \RBSL\ fields associated with the axial flux $F$, the corresponding values ${\bm A}^{*}_F(N,\xv)$, ${\bm B}^{*}_F(N,\xv)$, and ${\bm j}^{*}_F(N,\xv)$ must be added to rather than subtracted from the respective integrals over the coronal path $\mathcal{C}$, since the integrals over the subphotospheric path $\Cst$ are already taken with minus due to the assumed mirroring of these quantities (see Eqs. (\ref{AF}), (\ref{BF_int}), and (\ref{jF_int})).

As mentioned in Section \ref{s:fnu}, by representing our \RBSL\ integrals in terms of solutions of the above ODEs, we gain the advantage of being able to employ the existing developed machinery for solving ODEs with adaptive step and controlled accuracy of integration. The use of this machinery significantly facilitates the implementation of our method.

\section{Current Density due to Axial Magnetic Flux in a Toroidal \RBSL\ MFR 
\label{s:JFtorus}}

To reveal how ${\bm j}_{F}$ depends on the curvature of MFRs modeled with \RBSL{s}, let us derive its  expression for the particular case of toroidal MFRs. The axis path in this case is a circle, which is more conveniently parameterized by the arc angle $\vartheta$. By changing $\nu$ for $\vartheta$ one can rewrite  Eq. (\ref{jBdiff}) as follows:
\begin{eqnarray}
\mathrm{d}
\jF(\vartheta,\xv)
&=&
\frac{1}{a}
K_{j_F}(r)
\,
\mathrm{d}
\Rv
\times
\rv
\, , 
\qquad
\dbr
{
\frac
{F}
{4\pi\mu a^3}
}
\, .
	\label{djF}
\end{eqnarray}
Assume that the vector $\Rv$ here is defined in the system of coordinates with the origin at the center of symmetry of the torus. Using then the Frenet-Serret frame $(\Tv, \Nv, \Mv)$  at the point $\vartheta = 0$ of the path, we obtain
\begin{eqnarray}
\Rv
=
\frac{1}{\kappa}
\left(
\sin\vartheta
\,
\Tv
-
\cos\vartheta
\,
\Nv
\right)
\, ,
	\label{Rvtor}
\end{eqnarray}
where $\kappa$ is the curvature of the torus axis.

Due to the rotational symmetry of the torus, it is sufficient to derive $\jF$ in one cross-section of the torus, e.g., in the plane spanned on the above $\Nv$ and $\Mv$. Any point $\xv$ belonging to this plane is
\begin{eqnarray}
\xv
=
-
\frac{1}{\kappa}
\Nv
+
a
\ero
\equiv
-
\frac{1}{\kappa}
\Nv
+
a
\rho
\left(
\cos\omega
\,
\Nv
+
\sin\omega
\,
\Mv
\right)
\, ,
	\label{xvtor}
\end{eqnarray}
where $a \rho$ is the distance from the axis to  this point, and $\omega$ is the angle between $\ero$ and $\Nv$. Thus, $\rho$ and $\omega$ fully define $\xv$ in the plane $\vartheta=0$, so we will use them to determine the corresponding cross-sectional distribution of $\jF$.

Substituting Eqs. (\ref{Rvtor}) and (\ref{xvtor}) into $\rv = \left(\xv - \Rv \right)/a$, we obtain
\begin{eqnarray}
\rv
=
-
\frac
{
\sin
\vartheta
}
{
\kappa
a
}
\Tv
+
\left(
\rho
\,
\cos\omega
-
\frac
{
1
-
\cos\vartheta
}
{
\kappa
a
}
\right)
\Nv
+
\rho
\,
\sin\omega
\,
\Mv
\, ,
	\label{rvtor}
\end{eqnarray}
from where it follows that
\begin{eqnarray}
r^2
=
\rho^2
+
\frac
{2}
{
\kappa^2
a^2
}
\left(
1
-
\kappa
a
\rho
\,
\cos\omega
\right)
\left(
1
-
\cos\vartheta
\right)
\, .
	\label{r2tor}
\end{eqnarray}
One immediately deduces from here that $r$ is an even function of $\vartheta$ and
\begin{eqnarray}
\cos\vartheta
&=&
1
-
\frac
{
\kappa^2 a^2
\left(
r^2
-
\rho^2
\right)
}
{
2
\,
(
1
-
\kappa a
\rho\,\cos\omega
)
}
\, .
	\label{cosor}
\end{eqnarray}

With the help of Eqs. (\ref{Rvtor}) and (\ref{rvtor}), we can now rewrite Eq. (\ref{djF}) as follows:
\begin{eqnarray}
\mathrm{d}
\jF(\vartheta,\rho,\omega)
&=&
\frac{1}{a}
K_{j_F}(r)
\left[
\frac
{
\rho
\,
\sin\omega
}
{
\kappa
a
}
\sin\vartheta
\,
\Tv
+
\frac
{\rho}
{
\kappa
a
}
\cos\vartheta
\,
\eo
+
\frac
{1
-
\cos\vartheta
}
{
\kappa^2
a^2
}
\Mv
\right]
\mathrm{d}
\vartheta
\, ,
	\label{djFexp}
\end{eqnarray}
where
\begin{eqnarray}
\eo
\equiv
\ero
\times
\Tv
=
-
\sin\omega
\,
\Nv
+
\cos\omega
\,
\Mv
	\label{eo}
\end{eqnarray}
is a unit vector parallel to the azimuthal component of $\mathrm{d}\jF$ at the point $\xv$ defined by Eq. (\ref{xvtor}). The integration of Eq. (\ref{djFexp}) over $\vartheta$ from $-\pi$ to $\pi$ provides, in principle, the desired $\jF(\rho,\omega)$. Note, however, that the $\Tv$-component of Eq. (\ref{djFexp}) is an odd function of $\vartheta$, therefore the corresponding integral  vanishes identically. To integrate the remaining components of Eq. (\ref{djFexp}), it is convenient to change the  variable of integration from $\vartheta$ to $r$ with the help of Eq. (\ref{cosor}). This change of variables makes, in particular, the passage to the limit of vanishing $\kappa a $ in $\jF$  more transparent. Indeed, taking into account that $K_{j_F}(r) \equiv 0$ at $r>1$ (see Eq. (\ref{KjF})) and $\eo$- and $\Mv$-components of Eq. (\ref{djFexp}) are even functions of $\vartheta$, we obtain, after some calculations using Eq. (\ref{cosor}),
\begin{eqnarray}
\jF
&=&
j_{F\omega}
(\rho,\omega)
\,
\eo
+
j_{FM}
(\rho,\omega)
\,
\Mv
\, ,
\qquad
\dbr
{
\frac
{F}
{4\pi\mu a^3}
}
\, ,
	\label{jF_tor}
\\
j_{F\omega}
(\rho,\omega)
&=&
\frac
{
2
\rho
}
{
\left(
1
-
\kappa
a
\rho
\,
\cos\omega
\right)^{1/2}
}
\int^1_\rho
\frac
{\cos\vartheta}
{
\sqrt{
\left(
1
+
\cos\vartheta
\right)/2
}
}
\frac
{
r
\,
K_{j_F}(r)
\,
\mathrm{d}r
}
{
\sqrt{r^2-\rho^2}
}
\, ,
	\label{jFo_tor}
\\
j_{FM}
(\rho,\omega)
&=&
\frac
{
\kappa
a
}
{
\left(
1
-
\kappa
a
\rho
\,
\cos\omega
\right)^{3/2}
}
\int^1_\rho
\left[
\frac
{
2
\left(
r^2-\rho^2
\right)
}
{
1
+
\cos\vartheta
}
\right]
^{1/2}
r
\,
K_{j_F}(r)
\,
\mathrm{d}r
\, .
	\label{jFM_tor}
\end{eqnarray}
%
%
\begin{figure*}[ht!]
\plotone{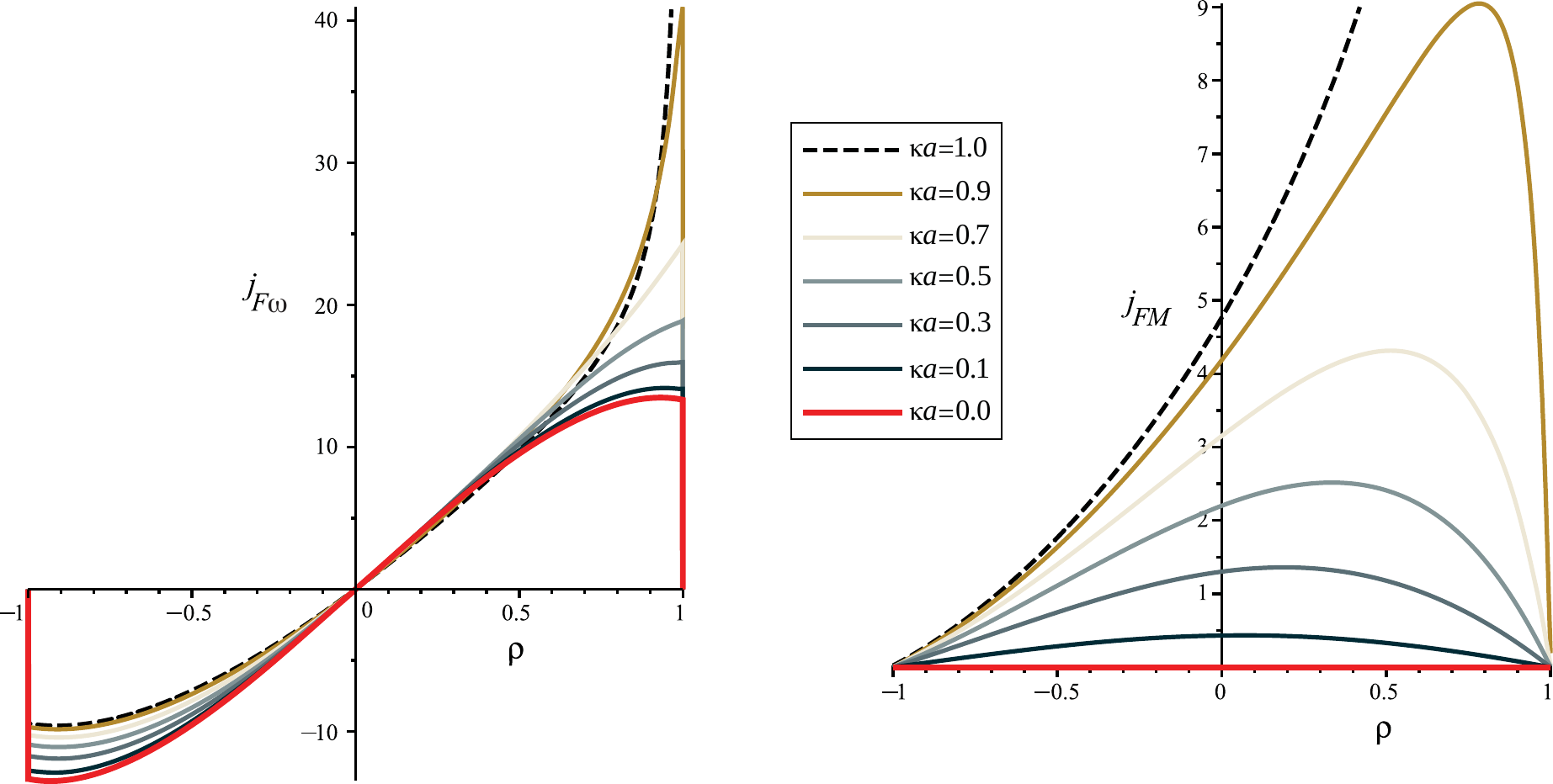}
\caption{
Distributions of the azimuthal ($j_{F\omega}$, Eq. (\ref{jFo_tor})) and binormal ($j_{FM}$, Eq. (\ref{jFM_tor})) components of the current density $\jF$ along the normal $\Nv$ to the torus axis at different values of the parameter $\kappa a$.
Both components are normalized to $F/(4\pi\mu a^3)$.
 	\label{jFoM}}
\end{figure*}
%
One can check that in the limit of vanishing curvature, $\kappa a \rightarrow 0$, the obtained component $j_{F\omega}(\rho,\omega) $ turns exactly into $j_\mathrm{az}(\rho)$ given by Eq. (\ref{jaz}), while $j_{FM}(\rho,\omega) $ vanishes identically, as required. Figure \ref{jFoM} displays the distributions of these components along the normal $\Nv$ at several other values of $\kappa a$, which here is identical with the inverse aspect ratio of the torus. In these plots, positive and negative radii $\rho$ simply correspond to the opposite directions of the radii defined by $\omega = 0$ and $\omega = \pi$, respectively. This signed $\rho$ coincides also with $y/a$, where $y$ is the coordinate axis mentioned previously in connection with Eq. (\ref{f}). The plots demonstrate that both components of $\jF$ have asymmetric profiles with larger amplitudes at $\rho>0$, i.e., toward the inner side of the torus. The asymmetry grows with $\kappa a$ and becomes particularly strong starting from $\kappa a \simeq 0.7$. At $\kappa a = 1$, it reaches its apogee when both components become singular at $\rho=1$, i.e., at the inner side of the torus (see dashed black curves in Figure \ref{jFoM}).

The considered effect is of a local nature, since the kernel $K_{j_F}(r)$ vanishes at $r>1$. In addition, toroidal MFRs locally approximate MFRs with other shapes and circular cross-sections. Therefore, the singular behavior of $\jF$ found for toroidal MFRs must also be present in all other  MFRs modeled with \RBSL{s}. Thus, when constructing these models, one needs to limit from above possible variations  of $\kappa a$ along the axis path to prevent the development of such singularities in the modeled MFRs, as discussed earlier in Section \ref{s:fnu}.


\bibliographystyle{aasjournal}
\bibliography{RBSL2}
\clearpage

\end{document}